\documentclass[aps,PRB,twocolumn,superscriptaddress,preprintnumbers]{revtex4-2}
\usepackage[T1]{fontenc}
\usepackage{times}
\usepackage{graphicx,color}
\usepackage{amsfonts,amsmath,amssymb,amsbsy}
\usepackage[colorlinks=true, linkcolor=blue, citecolor=blue, urlcolor=blue]{hyperref}
\usepackage{float}

\begin{document}

\title{Non-Clifford symmetry protected topological hyper-cluster states \\
and multi-qubit universal measurement-based quantum computation }
\author{Motohiko Ezawa}
\affiliation{Department of Applied Physics, The University of Tokyo, 7-3-1 Hongo, Tokyo
113-8656, Japan}

\begin{abstract}
A cluster state is a highly entangled quantum state that serves as a
universal resource for measurement-based quantum computation. It is
generated by applying controlled-Z (CZ) gates to the product state $\left\vert ++\cdots +\right\rangle $, and its parent Hamiltonian is the ZXZ
model. This model exhibits a topological phase protected by the $\mathbb{Z}_{2}^{\text{even}}\times \mathbb{Z}_{2}^{\text{odd}}$\ symmetry. By applying general quantum gates to
the state $\left\vert ++\cdots +\right\rangle $, we systematically obtain a
general short-range entangled cluster state. If we use the
controlled-controlled Z (CCZ) gate instead of the CZ gate, we obtain
hyper-cluster states featuring five-body interaction terms. These states enable the implementation of the CZ
gate within measurement-based quantum computation. Motivated by this, we propose a generalization to the  C$^{N}$Z gate, leading to a hyper-cluster model, in which $(2N+1)$-body entangled states are generated. This model retains the  $\mathbb{Z}_{2}^{\text{even}}\times
\mathbb{Z}_{2}^{\text{odd}}$\ symmetry, which lies outside the Clifford group for $N\geq
3$. We demonstrate that an open chain in this model exhibits a $2^{2N}$ fold degenerate ground states protected by the symmetry, corresponding to the emergence of $N$ effective free spins at each edge. These boundary degrees of freedom can serve as $N$ qubits for measurement-based quantum computation. In particular, this framework makes it possible to implement the C$^{N-1}$Z
gate in measurement-based quantum computation.
\end{abstract}

\date{\today }
\maketitle

\section{Introduction}

Quantum computation is a next generation computation paradigm grounded in the principles of quantum
mechanics\cite{Feynman,DiVi,NielsenC}. According to the Solovay-Kitaev
theorem, universal quantum computation\cite{Deutsch,Dawson,Universal} can be achieved using only $\pi /4$\ phase shift gate, the Hadamard
gate and the controlled-NOT (CNOT) gate. The standard approach is a gate-based
quantum computation\cite{Deutsch,Dawson,Universal}, in which a quantum circuit
is implemented by sequentially applying quantum gates. This method is efficient for
qubits realized in superconductors, trapped-ions and quantum dots. A serious
limitation, however, is their short coherent time, which makes it
difficult to apply a large number of quantum gates reliably. 

In contrast, photons possess extremely  long
coherent time, making photonic quantum computation a promising direction.
The challenge is that photons interact only very weakly, so implementing quantum gates directly is difficult. Measurement-based quantum computation (MBQC)%
\cite{Rauss,RaussA} offers a potential solution, because the unitary evolution of the logical state is achieved solely through a sequence of adaptive measurements. The procedure is as follows: one first prepares a cluster state, which is a highly entangled resource state. Then, by measuring selected qubits in appropriately chosen bases and applying feed-forward corrections, one realizes the desired unitary transformations.

We consider a chain made of $2L$ qubits. A one-dimensional cluster state $\left\vert \psi \right\rangle _{\text{CZ}}$ is commonly generated\cite%
{Briegel} by applying the controlled-Z (CZ) gates to the state $\bigotimes_{j=1}^{2L}\left\vert +\right\rangle =\left\vert ++\cdots
+\right\rangle $,%
\begin{equation}
\left\vert \psi \right\rangle _{\text{CZ}}=\prod_{j=1}^{2L-1}\text{CZ}%
_{j,j+1}\bigotimes_{j=1}^{2L}\left\vert +\right\rangle ,
\label{ClusterState}
\end{equation}%
where CZ$_{j,j+1}$ represents the CZ gate,\ whose control qubit is $j$-th
qubit and target qubit is ($j+1$)-th qubit.

The Hamiltonian which has the state $\left\vert \psi \right\rangle _{\text{CZ%
}}$ as the ground state is the ZXZ model\cite{Son,Nielsen},%
\begin{equation}
\mathcal{H}_{\text{ZXZ}}=-\sum_{j=2}^{2L-1}\text{Z}_{j-1}\text{X}_{j}\text{Z}%
_{j+1},  \label{HZXZ2}
\end{equation}%
where X$_{j}$ (Z$_{j}$) represents the Pauli gate $\sigma _{x}$ ($\sigma _{z}
$) acting on the $j$-th qubit. It is a simplest example realizing the
symmetry-protected topological (SPT) phase\cite{Gu,Poll} protected by the $%
\mathbb{Z}_{2}^{\text{even}}\times \mathbb{Z}_{2}^{\text{odd}}$ symmetry.
The ground state is gapped and unique for a closed chain. It is a
short-ranged entangled state\cite{SRE}, which is disentangled by a
finite-depth local-unitary quantum circuit. On the other hand, the ground
state is four-fold degenerate for an open chain. They are used as 2 qubits,
which are robust against decoherence preserving the $\ \mathbb{Z}_{2}^{\text{%
even}}\times \mathbb{Z}_{2}^{\text{odd}}$ symmetry\cite{Paszko}. They form
logical qubits in MBQC\cite{Rauss,RaussA,Else12}. For example, the left
qubit acts as an input of information. The quantum computation is done by
measuring bulk qubits. Then, the output is readout from the right qubit.
Recently, it is recognized\cite{Mana,Seif} that this model is a good example
possessing non-invertible symmetry\cite{Seiberg,Seiberg2}. The cluster state
can be experimentally realized by using entangled photons\cite{Schwartz}.
The ZXZ model is simulated in quantum computer by employing the method of
quantum imaginary-time evolution\cite{CHLee}. The standard MBQC is based on
the CZ cluster state, where we perform a measurement of the qubit in the
basis given by%
\begin{equation}
\lvert \alpha _{s}(\theta )\rangle =\frac{1}{\sqrt{2}}\bigl(\lvert 0\rangle
+(-1)^{s}e^{i\theta }\lvert 1\rangle \bigr),\quad s\in \{0,1\}.
\end{equation}%
The one-qubit universal quantum gate is implemented based on this
single-qubit measurement. A network structure is used to implement the CZ
gate in the standard MBQC.

In this paper, by using a general quantum gate, we generalize a short-ranged
entangled cluster state (\ref{ClusterState}) as 
\begin{equation}
\left\vert \psi \right\rangle _{U}=\prod_{j=1}^{2L-N}U_{\left\{ j;N\right\}
}\bigotimes_{j=1}^{2L}\left\vert +\right\rangle ,  \label{psiU}
\end{equation}%
where $U_{\left\{ j;N\right\} }$ is a finite-depth local-unitary $N$-qubit
gate around the $j$-th qubit. Then, we construct a cluster Hamiltonian
associated with the cluster state (\ref{psiU}), which gives a generalized
ZXZ model. If we use the C$^{N}$Z gate, we obtain a hyper-cluster models
with a $\left( 2N+1\right) $-body interaction. It is protected by
non-Clifford symmetry for $N\geq 3$. We show that the CZ gate is implemented
by measuring one qubit with angle $\pi /2$, which is an indispensable
element for universal quantum computation. Hence, it is possible to execute
universal quantum computation. The $\mathbb{Z}_{2}^{\text{even}}\times \mathbb{Z}_{2}^{\text{%
odd}}$ symmetry and non-invertible symmetry are obtained by a corresponding
unitary transformation. It is remarkable that there are $2^{2N}$ fold
degenerate $\mathbb{Z}_{2}^{\text{even}}\times \mathbb{Z}_{2}^{\text{odd}}$ symmetry-protected
ground states for an open chain, which indicates that $N$ free spins emerge
at each edge. They can be used as $N$ qubits in MBQC. Especially, the C$%
^{N-1}$Z gate is directly implemented based on the C$^{N}$Z cluster state,
which is difficult in general quantum computation.

This paper is composed as follows. In Sec.II, we briefly summarize the
results on the ZXZ model and study topological phase transition between the
ZXZ model and the trivial X model. In Sec.III, we introduce a general
cluster state based on an arbitrary unitary gate. In Sec.IV, we study a
hyper-cluster model generated by the CCZ gate, which has five-body
interactions of qubits. We derive that the CZ gate is implemented by
measuring one qubit in the CCZ cluster state, and hence, it is possible to
execute universal MBQC. In Sec.V, we study a cluster model generated by the C%
$^{N}$Z gates with $1\leq n\leq N$. It is shown that the model has $\left(
2N+1\right) $-body interaction. There emerge $2^{2N}$\ fold degenerate
ground states for an open chain, which indicates that $N$\ free spins emerge
at each edge. They can be used as $N$ qubits in MBQC. We show that the C$%
^{N-1}$Z gate is directly implemented based on the C$^{N}$Z cluster state by
measuring one qubit.  Sec.VI is devoted to discussions.

\section{ZXZ model}

We first review the cluster state and the ZXZ model\cite{Son,Nielsen}. We
consider a chain with the length $2L$ where $L\geq 1$. The cluster state is
generated by applying the CZ gates to the state $\bigotimes_{j=1}^{2L}\left%
\vert +\right\rangle $ as%
\begin{equation}
\left\vert \psi ^{\text{ZXZ}}\right\rangle =\text{CZ}\bigotimes_{j=1}^{2L}%
\left\vert +\right\rangle ,  \label{psiZXZ}
\end{equation}%
where%
\begin{equation}
\text{CZ}\equiv \prod_{j=1}^{2L-1}\text{CZ}_{j,j+1}
\end{equation}%
for an open chain, while 
\begin{equation}
\text{CZ}\equiv \prod_{j=1}^{2L}\text{CZ}_{j,j+1}=\text{CZ}%
_{2L,1}\prod_{j=1}^{2L-1}\text{CZ}_{j,j+1}
\end{equation}%
for a closed chain by identifying the ($2L+1$)-th qubit with the first
qubit. Here,%
\begin{align}
\text{CZ}_{j,j+1}\equiv & 1-\frac{1-\text{Z}_{j}}{2}\frac{1-\text{Z}_{j+1}}{2%
}  \notag \\
=& \text{diag}\left( 1,1,1,-1\right) _{j,j+1}
\end{align}%
is the CZ gate acting on the $j$-th and ($j+1$)-th qubits.

\begin{figure}[t]
\centerline{\includegraphics[width=0.48\textwidth]{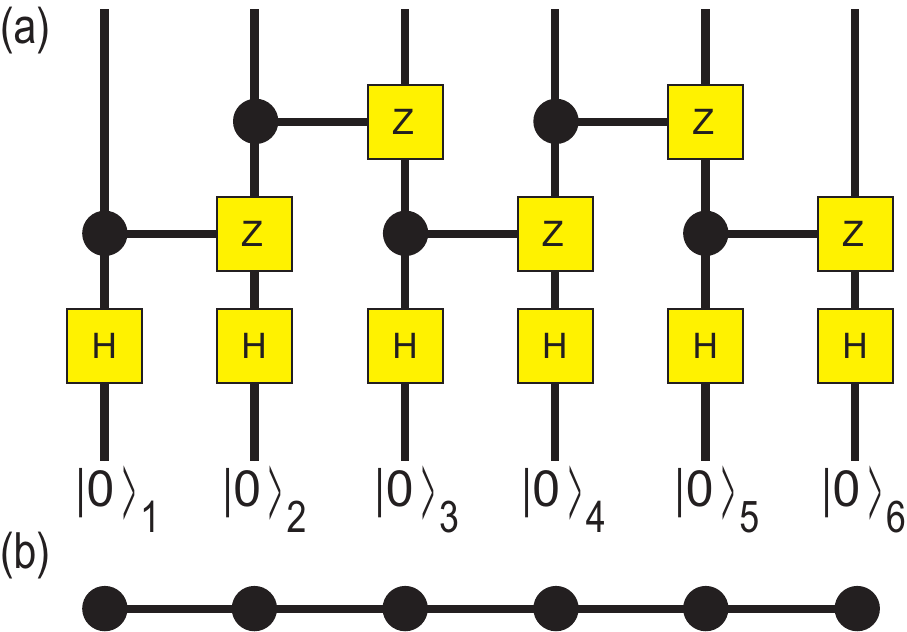}}
\caption{ZXZ model. (a) Quantum circuit generating the ZXZ cluster state.
(b) Its graph representation, where the black disks represent the state $%
\left\vert +\right\rangle $ and the black segments represent the CZ gates.}
\label{FigZXZ}
\end{figure}

The state $\bigotimes_{j=1}^{2L}\left\vert +\right\rangle $ is generated by
the Hadamard gate as%
\begin{equation}
\bigotimes_{j=1}^{2L}\left\vert +\right\rangle =\bigotimes_{j=1}^{2L}\text{H}%
_{j}\left\vert 0\right\rangle ,
\end{equation}%
where $\left\vert +\right\rangle =\left( \left\vert 0\right\rangle
+\left\vert 1\right\rangle \right) /\sqrt{2}$, and 
\begin{equation}
\text{H}_{j}\equiv \frac{1}{\sqrt{2}}\left( 
\begin{array}{cc}
1 & 1 \\ 
1 & -1%
\end{array}%
\right) 
\end{equation}%
is the Hadamard gate acting on the $j$-th qubit. The quantum circuit
corresponding to Eq.(\ref{psiZXZ}) is shown in Fig.\ref{FigZXZ}(a). The
generated cluster state is called the graph state\cite{RaussA,Hein}, because
it can be fully specified by gates acting on the edges of a graph. The graph
is undirected because CZ gates are symmetric between target and control, and
because all CZ gates commute one another,%
\begin{equation}
\left[ \text{CZ}_{j,j+1},\text{CZ}_{j^{\prime },j^{\prime }+1}\right] =0.
\end{equation}%
The graph representation of Eq.(\ref{psiZXZ}) is shown in Fig.\ref{FigZXZ}%
(b).

The corresponding cluster Hamiltonian is composed as follows. We start with
the Hamiltonian 
\begin{equation}
\mathcal{H}_{\text{X}}=-\sum_{j=1}^{2L}\text{X}_{j},  \label{HX}
\end{equation}%
whose ground state is $\bigotimes_{j=1}^{2L}\left\vert +\right\rangle $
because we have X$_{j}\left\vert +\right\rangle =\left\vert +\right\rangle ,$
where X$_{j}$ is the Pauli X operator acting on the $j$-th qubit. Then, the
Hamiltonian 
\begin{equation}
\mathcal{H}_{\text{ZXZ}}\equiv \text{CZ}\mathcal{H}_{\text{X}}\text{CZ}%
=-\sum_{j=1}^{2L}\left( \text{CZ}\right) \text{X}_{j}\left( \text{CZ}\right) 
\label{HZXZCZ}
\end{equation}%
has the ground state $\left\vert \psi ^{\text{ZXZ}}\right\rangle $ given by (%
\ref{psiZXZ}). By using the relations%
\begin{align}
\text{CZ}_{j,j+1}\text{X}_{j}\text{CZ}_{j,j+1}=& \text{X}_{j}\text{Z}_{j+1},
\\
\text{CZ}_{j,j+1}\text{X}_{j+1}\text{CZ}_{j,j+1}=& \text{Z}_{j}\text{X}%
_{j+1},
\end{align}%
we have%
\begin{align}
& \left( \text{CZ}\right) \text{X}_{j}\left( \text{CZ}\right)   \notag \\
=& \left( \text{CZ}_{j-1,j}\right) \left( \text{CZ}_{j,j+1}\right) \text{X}%
_{j}\left( \text{CZ}_{j,j+1}\right) \left( \text{CZ}_{j-1,j}\right)   \notag
\\
=& \left( \text{CZ}_{j-1,j}\right) \text{X}_{j}\text{Z}_{j+1}\left( \text{CZ}%
_{j-1,j}\right)   \notag \\
=& \left( \text{CZ}_{j-1,j}\right) \text{X}_{j}\left( \text{CZ}%
_{j-1,j}\right) \text{Z}_{j+1}  \notag \\
=& \text{Z}_{j-1}\text{X}_{j}\text{Z}_{j+1}
\end{align}%
for the bulk $j$-th qubit with $1\leq j\leq N-1$.

\begin{figure}[t]
\centerline{\includegraphics[width=0.48\textwidth]{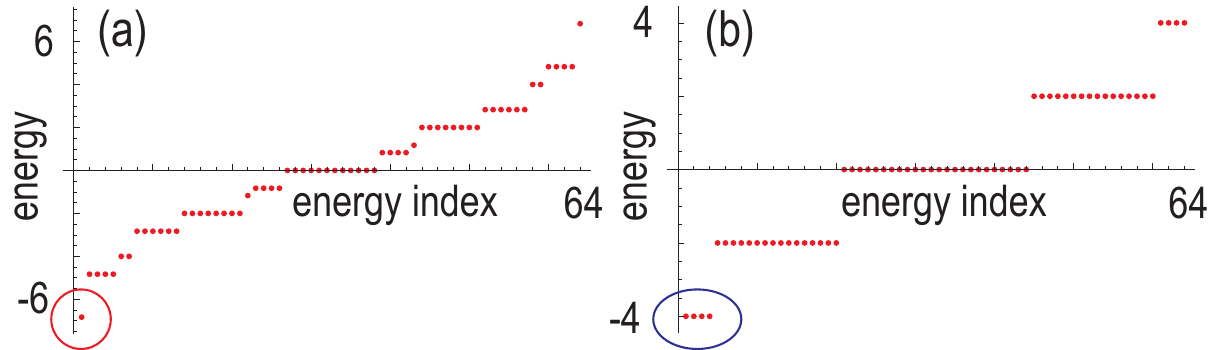}}
\caption{Energy spectrum of the ZXZ model. (a) Closed chain, where there is
a single gapped ground state as indicated by a closed red circle. (b) Open
chain, where there are four-fold degenerate gapped ground state as indicated
by a closed blue oval. We have set $L=3$. The vertical axis is the energy,
while the horizontal axis is the index of the energy.}
\label{FigZXZEne}
\end{figure}

Then, the Hamiltonian (\ref{HZXZCZ}) is given in the form of%
\begin{equation}
\mathcal{H}_{\text{ZXZ}}=-\sum_{j=1}^{2L}K_{j}^{\text{ZXZ}},
\label{HZXZOpen}
\end{equation}%
with 
\begin{equation}
K_{j}^{\text{ZXZ}}\equiv \text{Z}_{j-1}\text{X}_{j}\text{Z}_{j+1}.
\label{KZXZj}
\end{equation}%
The ground state $\left\vert g\right\rangle $ satisfies%
\begin{equation}
K_{j}^{\text{ZXZ}}\left\vert g\right\rangle =\left\vert g\right\rangle 
\label{stabilizerg}
\end{equation}%
for all $j$, where $K_{j}^{\text{ZXZ}}$ is called the stabilizer and $%
\left\vert g\right\rangle $ is called the stabilizer state. We note that the
explicit forms of the stabilizers at edges $j=1$ and $N$ depend on an open
chain or a closed chain. See Appendix \ref{ApZXZ}.

By using Eq.(\ref{KZXZj}), the stabilizers satisfy%
\begin{equation}
\left[ K_{j}^{\text{ZXZ}},K_{j^{\prime }}^{\text{ZXZ}}\right] =0,\qquad
\left( K_{j}^{\text{ZXZ}}\right) ^{2}=1.
\end{equation}%
The Hamiltonian (\ref{HZXZOpen}) is frustration free\cite{AKLT1,AKLT2,Kitaev}%
, where the ground state is exactly solvable even though it is a strongly
correlated Hamiltonian. The energy spectra are shown in Fig.\ref{FigZXZEne}
for closed and open chains. The stabilizers of the closed and open chains,
symmetry properties and string-order parameter are summarized in Appendix %
\ref{ApZXZ}.

\subsubsection{$\mathbb{Z}_{2}^{\text{even}}\times \mathbb{Z}_{2}^{\text{odd}%
}$ symmetry}

The ZXZ Hamiltonian $\mathcal{H}_{\text{ZXZ}}$ satisfies%
\begin{equation}
\left[ \mathcal{H}_{\text{ZXZ}},\eta _{\text{ZXZ}}^{\text{even}}\right] =%
\left[ \mathcal{H}_{\text{ZXZ}},\eta _{\text{ZXZ}}^{\text{odd}}\right] =0,
\end{equation}%
with%
\begin{equation}
\eta _{\text{ZXZ}}^{\text{even}}\equiv \prod\limits_{j\in \text{even}}\text{%
X}_{j},\qquad \eta _{\text{ZXZ}}^{\text{odd}}\equiv \prod\limits_{j\in 
\text{odd}}\text{X}_{j}.  \label{etaZXZ}
\end{equation}%
Because of the relations%
\begin{equation}
\left( \eta _{\text{ZXZ}}^{\text{even}}\right) ^{2}=\left( \eta _{\text{ZXZ}%
}^{\text{odd}}\right) ^{2}=1,
\end{equation}%
the Hamiltonian $\mathcal{H}_{\text{ZXZ}}$\ given by Eq.(\ref{HZXZK}) has
the $\mathbb{Z}_{2}^{\text{even}}\times \mathbb{Z}_{2}^{\text{odd}}$
symmetry. Consequently, it defines a one-dimensional SPT phase classified by
the second cohomology group\cite{Chen,Else} according to 
\begin{equation}
H^{2}\left( \mathbb{Z}_{2}^{\text{even}}\times \mathbb{Z}_{2}^{\text{odd}%
},U\left( 1\right) \right) =\mathbb{Z}_{2},
\end{equation}%
where $U\left( 1\right) $ is the group associated with the projective
representation of the edge states. See Appendix \ref{ApZXZco} for the second
cohomology group.

\begin{figure}[t]
\centerline{\includegraphics[width=0.48\textwidth]{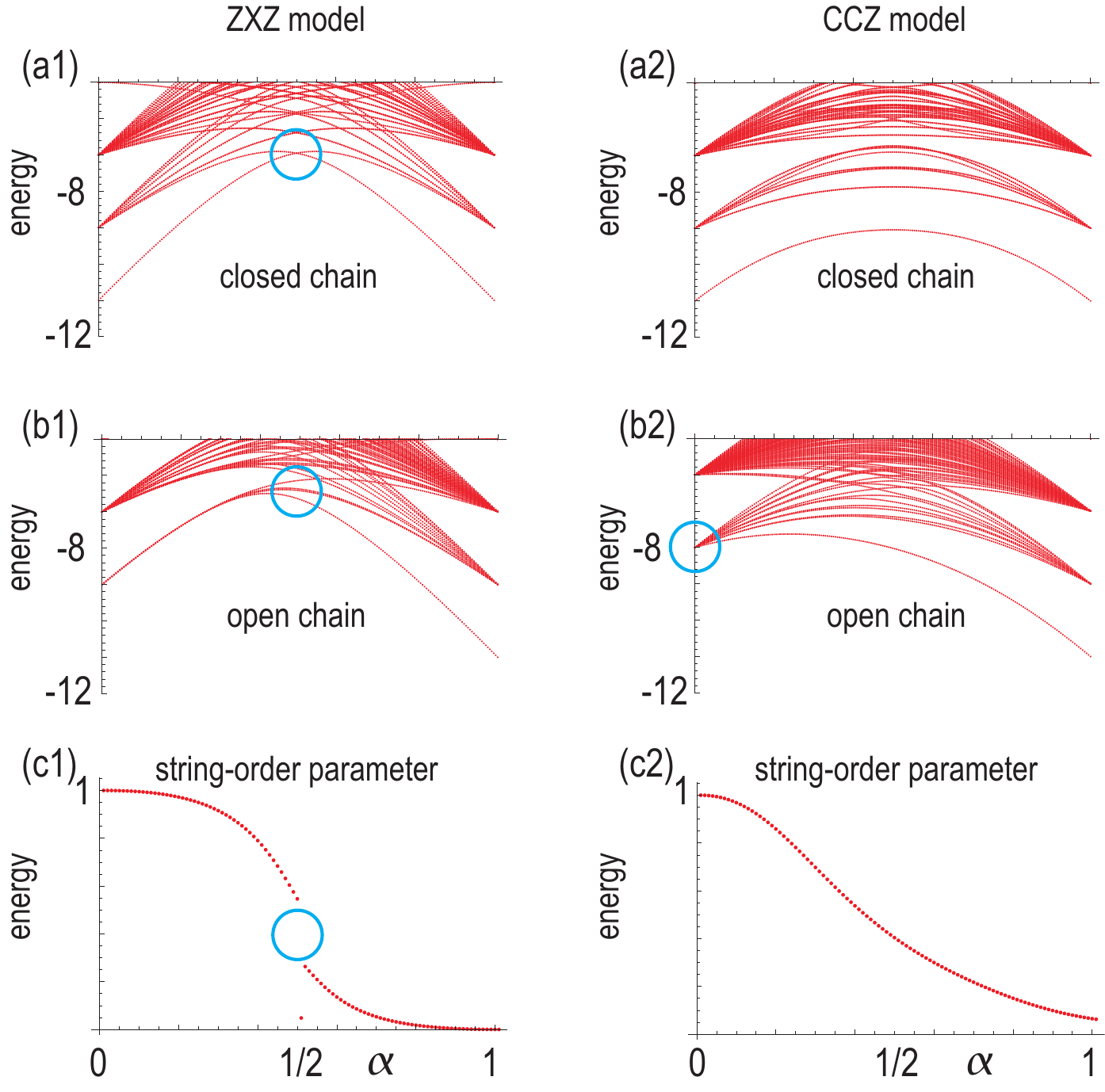}}
\caption{(a1) Energy spectrum of the ZXZ model for a closed chain. The
energy is symmetric at $\protect\alpha =1/2$ reflecting the $\mathbb{Z}_{2}^{%
\text{CZ}}$ symmetry. (b1) That for an open chain. The ground states are
four-fold degenerate at $\protect\alpha =0$ and not degenerate at $\protect%
\alpha =1$. The degeneracy splits at $\protect\alpha $ slightly smaller than 
$1/2$. (c1) String order parameter. There is a jump at $\protect\alpha =1/2$%
, implying a topological phase transition at $\protect\alpha =1/2$. (a2)
Energy spectrum of the CCZ model for a closed chain. (b2) That for an open
chain. The ground states are 16-fold degenerate at $\protect\alpha =0$ and
not degenerate at $\protect\alpha =1$. The degeneracy splits for $\protect%
\alpha >0$. (c2) String order parameter for the CCZ model. There is no jump
at $\protect\alpha =1/2$, implying a topological phase transition at $%
\protect\alpha =0$. The horizontal axis is $\protect\alpha $. We have set $%
L=11$}
\label{FigTra}
\end{figure}

\subsubsection{Topological phase transition}

We consider a Hamiltonian interpolating $\mathcal{H}_{\text{ZXZ}}$ and $%
\mathcal{H}_{\text{X}}$,%
\begin{equation}
H\left( \alpha \right) =\left( 1-\alpha \right) \mathcal{H}_{\text{ZXZ}%
}+\alpha \mathcal{H}_{\text{X}}
\end{equation}%
with $0\leq \alpha \leq 1$. It has a duality generated by the CZ gate,%
\begin{equation}
\text{CZ}\mathcal{H}\left( \alpha \right) \text{CZ}=\mathcal{H}\left(
1-\alpha \right) ,  \label{CZdual}
\end{equation}%
where it is self dual at $\alpha =1/2$. There emerges an enhanced $\mathbb{Z}%
_{2}^{\text{even}}\times \mathbb{Z}_{2}^{\text{odd}}\times \mathbb{Z}_{2}^{%
\text{CZ}}$ symmetry at $\alpha =1/2$. The energy spectrum for a closed
chain is shown as a function of $\alpha $ in Fig.\ref{FigTra}(a1). The
energy spectrum is symmetric about $\alpha =1/2$, showing the duality
relation. The gap between the ground state and the first-excited state
closes at $\alpha =1/2$, indicating a topological phase transition. On the
other hand, the energy spectrum becomes asymmetric for an open chain as
shown in Fig.\ref{FigTra}(b1) because $\mathbb{Z}_{2}^{\text{CZ}}$ symmetry
is broken at the edges and the ground state degeneracies are different
between $\mathcal{H}_{\text{ZXZ}}$ and $\mathcal{H}_{\text{X}}$. The
four-fold degenerate edge states do not split for $\alpha <1/2$ because
there is only one spin at one edge. The degeneracy is broken at $\alpha $
slightly smaller than $1/2$, whose reason is a finite-size effect. The
degeneracy is broken at $\alpha =1/2$ for a sufficiently long chain. The
string order parameter is shown in Fig.\ref{FigTra}(c1) as a function of $%
\alpha $. It monotonically decreases as the increase of $\alpha $ and jumps
at $\alpha =1/2$, reflecting a topological phase transition.

Topological phase transition occurs at $\alpha =1/2$. The mechanism is the
same as in the Kramers-Wannier duality in the Ising model. There are only
two topological distinct phases because the second group cohomology is $%
\mathbb{Z}_{2}$. Then, a topological phase transition occurs only once. On
the other hand, $H\left( \alpha \right) $ and $H\left( 1-\alpha \right) $
are connected by the duality transformation as in Eq.(\ref{CZdual}). Hence,
the topological phase transition occurs at $\alpha =1/2$.

\section{Unitary-mapped cluster model}

We generalize the ZXZ model (\ref{HZXZ2}) by considering a cluster state
generated by%
\begin{equation}
\left\vert \psi \right\rangle _{U}=\prod_{j=1}^{2L-N}U_{\left\{ j;N\right\}
}\bigotimes_{j=1}^{2L}\left\vert +\right\rangle ,  \label{UjN}
\end{equation}%
where $U_{\left\{ j;N\right\} }$ is a finite-depth $N$-qubit gate around the 
$j$-th qubit instead of the CZ gate, and $\left\{ j;N\right\} \equiv \left\{
j,j+1,\cdots ,j+N\right\} $ represents the $N$ sequential numbers starting
from $j$. Explicit examples are given in Eqs.(\ref{psiCCZ}) and (\ref{psiCNZ}%
). The corresponding cluster Hamiltonian is constructed as%
\begin{equation}
\mathcal{H}_{U}=-\sum_{j}K_{j}
\end{equation}%
with%
\begin{equation}
K_{j}\equiv U_{\left\{ j;N\right\} }\text{X}_{j}U_{\left\{ j;N\right\}
}^{-1}.
\end{equation}%
It is related to the ZXZ Hamiltonian as%
\begin{equation}
\mathcal{H}_{U}=-VH_{\text{ZXZ}}V^{-1},
\end{equation}%
where%
\begin{equation}
V\equiv \prod_{j=1}^{2L-N}V_{\left\{ j;N\right\} }
\end{equation}%
with%
\begin{equation}
V_{\left\{ j;N\right\} }\equiv U_{\left\{ j;N\right\} }\text{CZ}%
_{j,j+1}^{-1}.
\end{equation}

The stabilizer state $\left\vert g_{U}\right\rangle $\ after the unitary
transformation is given by 
\begin{equation}
\left\vert g_{U}\right\rangle \equiv V\left\vert g\right\rangle 
\end{equation}%
for the Hamiltonian $H_{U}$, satisfying%
\begin{equation}
K_{j}^{U}\left\vert g_{U}\right\rangle =\left\vert g_{U}\right\rangle 
\end{equation}%
with%
\begin{equation}
K_{j}^{U}\equiv VK_{j}V^{-1}.
\end{equation}%
We note that the original stabilizer state is defined by the Pauli group as
in Eq.(\ref{KZXZj}).

\begin{figure}[t]
\centerline{\includegraphics[width=0.48\textwidth]{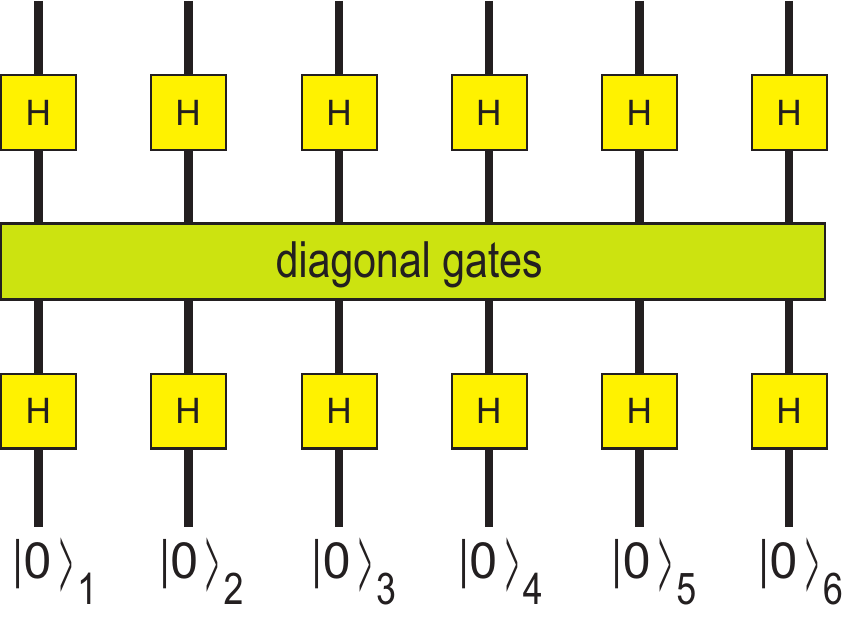}}
\caption{Quantum circuit for the class of the instantaneous quantum
polynomial time (IQP).}
\label{FigIQP}
\end{figure}

We mainly take unitary transformations made by a quantum circuit composed
only of diagonal gates and Hadamard gates as shown in Fig.\ref{FigIQP}. This
class of quantum circuits is known as the instantaneous quantum polynomial
time (IQP) class in the context of quantum information theory\cite{Brem}. It
does not present a universal computation but it is impossible to simulate a
quantum circuit in a polynomial time by a classical computer. We study
various models by taking various unitary operators $U_{\left\{ j;N\right\} }$%
, which are shown in Appendix \ref{ApXZX}$\sim $\ref{ApCNP}. Non-invertible
symmetry for general unitary-mapped cluster model is shown in Appendix \ref%
{ApNI}.

\section{CCZ gate model}

We study the cluster model generated by the CCZ gate, which is defined by%
\begin{align}
\text{CCZ}_{j-1,j,j+1}\equiv & 1-\frac{1-\text{Z}_{j-1}}{2}\frac{1-\text{Z}%
_{j}}{2}\frac{1-\text{Z}_{j+1}}{2}  \notag \\
=& \text{diag}\left( 1,1,1,1,1,1,1,-1\right) _{j-1,j,j+1},
\end{align}%
where ($j-1$)-th and $j$-th qubits are the controlled qubits and ($j+1$)-th
qubit is the target qubit. The action of the CCZ gate is%
\begin{equation}
\text{CCZ}_{0,1,2}\left\vert x_{0}x_{1}x_{2}\right\rangle =\left( -1\right)
^{x_{0}x_{1}x_{2}}\left\vert x_{0}x_{1}x_{2}\right\rangle 
\end{equation}%
with $x_{j}=0,1$ for $j=0,1,2$.

\begin{figure}[t]
\centerline{\includegraphics[width=0.48\textwidth]{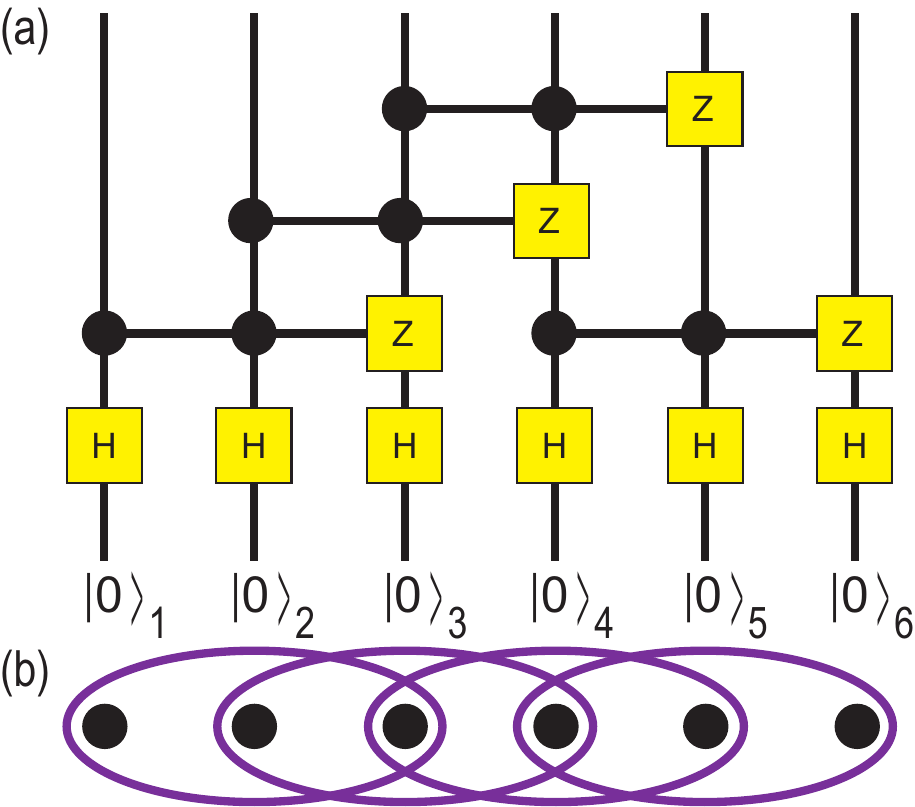}}
\caption{CCZmodel. (a) Quantum circuit representation. (b) Hypergraph
representation. Ovals including three vertices represent CCZ gates.}
\label{FigCCZCircuit}
\end{figure}

The CCZ gate is not a Clifford gate because it belongs to the Clifford
hierarchy with the level 3. See Appendix \ref{ApCl} for the Clifford
hierarchy. We consider a hyper-cluster state generated by the CCZ gate,%
\begin{equation}
\left\vert \psi \right\rangle _{\text{CCZ}}=\prod_{j=2}^{L-1}\text{CCZ}%
_{j-1,j,j+1}\bigotimes_{j=1}^{L}\left\vert +\right\rangle ,  \label{psiCCZ}
\end{equation}%
which is of the form of Eq.(\ref{UjN}) with $N=2$. We call it the CCZ
hyper-cluster state because it is generated by the hypergraph state\cite%
{Rossi,Huang} made of the CCZ gate. The quantum circuit corresponding to Eq.(%
\ref{psiCCZ}) is shown in Fig.\ref{FigCCZCircuit}(a). The hyper-graph
representation of Eq.(\ref{psiCCZ}) is shown in Fig.\ref{FigCCZCircuit}(b).

The CCZ gate is decomposed as%
\begin{align}
& \text{CCZ}_{j-1,j,j+1}  \notag \\
=& e^{i\frac{\pi }{8}}e^{-i\frac{\pi }{8}\text{Z}_{j-1}\otimes \text{I}%
_{j}\otimes \text{I}_{j+1}}e^{-i\frac{\pi }{8}\text{I}_{j-1}\otimes \text{Z}%
_{j}\otimes \text{I}_{j+1}}e^{-i\frac{\pi }{8}\text{I}_{j-1}\otimes \text{I}%
_{j}\otimes \text{Z}_{j+1}}  \notag \\
& \times e^{i\frac{\pi }{8}\text{Z}_{j-1}\otimes \text{Z}_{j}\otimes \text{I}%
_{j+1}}e^{i\frac{\pi }{8}\text{Z}_{j-1}\otimes \text{I}_{j}\otimes \text{Z}%
_{j+1}}e^{i\frac{\pi }{8}\text{I}_{j-1}\otimes \text{Z}_{j}\otimes \text{Z}%
_{j+1}}  \notag \\
& \times e^{-i\frac{\pi }{8}\text{Z}_{j-1}\otimes \text{Z}_{j}\otimes \text{Z%
}_{j+1}}.
\end{align}%
The product of the CCZ gates is given by%
\begin{align}
& \prod\limits_{j=1}^{2L}\text{CCZ}_{j-1,j,j+1}  \notag \\
=& e^{i\frac{\pi L}{8}}\prod\limits_{j=1}^{2L}e^{-i\frac{3\pi }{8}\text{Z}%
_{j}}e^{i\frac{\pi }{4}\text{Z}_{j}\otimes \text{Z}_{j+1}}e^{-i\frac{\pi }{8}%
\text{Z}_{j-1}\otimes \text{Z}_{j}\otimes \text{Z}_{j+1}}.  \label{CCZe}
\end{align}

Then, the corresponding Hamiltonian has a form%
\begin{equation}
H_{\text{CCZ}}=-\sum_{j}K_{j}^{\text{CCZ}}
\end{equation}%
with the stabilizer%
\begin{equation}
K_{j}^{\text{CCZ}}=\text{CCZ}_{\left\{ j\right\} }\text{X}_{j}\text{CCZ}%
_{\left\{ j\right\} },
\end{equation}%
where%
\begin{equation}
\text{CCZ}_{\left\{ j\right\} }\equiv \text{CCZ}_{j-2,j-1,j}\text{CCZ}%
_{j-1,j,j+1}\text{CCZ}_{j,j+1,j+2}.
\end{equation}

Especially, the Hamiltonian corresponding to the CCZ gate is given by%
\begin{align}
& H_{\text{CCZ}}\equiv \sum_{j=1}^{2L}K_{j}^{\text{CCZ}}  \notag \\
=& \frac{1}{4}\sum_{j=1}^{2L}\text{X}_{j}+\frac{1}{4}\sum_{j=1}^{2L-1}\text{X%
}_{j}\text{Z}_{j+1}+\text{Z}_{j}\text{X}_{j+1}  \notag \\
& +\frac{1}{4}\sum_{j=1}^{2L-2}\text{X}_{j}\text{Z}_{j+2}+\text{Z}_{j}\text{X%
}_{j+2}  \notag \\
& +\frac{1}{4}\sum_{j=1}^{2L-2}\text{Z}_{j}\text{X}_{j+1}\text{Z}_{j+2}+%
\text{X}_{j}\text{Z}_{j+1}\text{Z}_{j+2}+\text{Z}_{j}\text{Z}_{j+1}\text{X}%
_{j+2}  \notag \\
& -\frac{1}{4}\sum_{j=1}^{2L-3}\text{Z}_{j}\text{X}_{j+1}\text{Z}_{j+3}+%
\text{Z}_{j}\text{X}_{j+2}\text{Z}_{j+3}  \notag \\
& +\frac{1}{4}\sum_{j=1}^{2L-4}\text{Z}_{j}\text{X}_{j+2}\text{Z}_{j+4} 
\notag \\
& -\frac{1}{4}\sum_{j=1}^{2L-3}\text{Z}_{j}\text{Z}_{j+1}\text{X}_{j+2}\text{%
Z}_{j+3}+\text{Z}_{j}\text{X}_{j+1}\text{Z}_{j+2}\text{Z}_{j+3}  \notag \\
& +\sum_{j=1}^{2L-3}\text{Z}_{j}\text{Z}_{j+1}\text{X}_{j+2}\text{Z}_{j+4}+%
\text{Z}_{j}\text{X}_{j+2}\text{Z}_{j+3}\text{Z}_{j+4}  \notag \\
& -\frac{1}{4}\sum_{j=1}^{2L-4}\text{Z}_{j}\text{Z}_{j+1}\text{X}_{j+2}\text{%
Z}_{j+3}\text{Z}_{j+4},  \label{HCCZ}
\end{align}%
where we have used the relation%
\begin{equation}
\text{CCZ}_{abc}\text{X}_{a}\text{CCZ}_{abc}=\text{X}_{a}\text{CZ}_{bc}.
\label{CCZX}
\end{equation}%
It is summarized in the form of%
\begin{equation}
K_{j}^{\text{CCZ}}=\sum_{\left\{ \zeta _{j},\upsilon _{j}\right\} }\alpha
_{\zeta _{j-2}\zeta _{j-1}\zeta _{j+1}\zeta _{j+2}}\text{Z}_{j-2}^{\zeta
_{j-2}}\text{Z}_{j-1}^{\zeta _{j-1}}\text{X}_{j}\text{Z}_{j+1}^{\zeta _{j+1}}%
\text{Z}_{j+2}^{\zeta _{j+2}},
\end{equation}%
where $\zeta _{j},=0,1$ and%
\begin{align}
\alpha _{0000}=& \alpha _{0010}=\alpha _{0100}=\alpha _{0001}=\alpha _{1000}
\notag \\
=& \alpha _{0110}=\alpha _{0011}=\alpha _{1100}=\alpha _{1001}  \notag \\
=& \alpha _{1101}=\alpha _{1011}=1/4, \\
\alpha _{1110}=& \alpha _{0111}=\alpha _{1111}=-1/4.
\end{align}%
We note that the Hamiltonian (\ref{HCCZ}) is different from the previous
study on a hyper-cluster model with five-body spin interactions\cite%
{Minami,Yanagi,Seif,Fech,Cao,Li,Verr}.

The generators of the $\mathbb{Z}_{2}^{\text{even}}\times \mathbb{Z}_{2}^{%
\text{odd}}$ symmetry are given by%
\begin{align}
\eta _{\text{CCZ}}^{\text{even}}=& \prod\limits_{j\in \text{even}}\text{CZ}%
_{j,j+2}\prod\limits_{j\in \text{even}}\text{X}_{j}, \\
\eta _{\text{CCZ}}^{\text{odd}}=& \prod\limits_{j\in \text{odd}}\text{CZ}%
_{j,j+2}\prod\limits_{j\in \text{odd}}\text{X}_{j},
\end{align}%
where we have used the relation (\ref{CCZX}).

\begin{figure}[t]
\centerline{\includegraphics[width=0.48\textwidth]{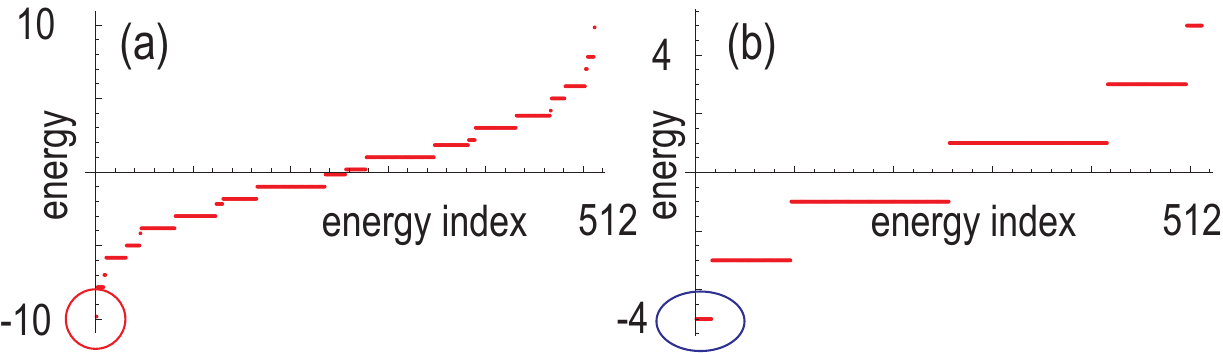}}
\caption{Energy spectrum of the CCZ model. (a) Closed chain, where there is
a single gapped ground state as indicated by a closed red circle. (b) Open
chain, where there are $16$-fold degenerate gapped ground states as
indicated by a closed blue oval. We have set $L=9$. The vertical axis is the
energy, while the horizontal axis is the index of the energy.}
\label{FigCCZEne}
\end{figure}

\subsection{Edge states}

We consider the $\mathbb{Z}_{2}^{\text{even}}\times \mathbb{Z}_{2}^{\text{odd%
}}$ symmetry projected Hamiltonian,%
\begin{equation}
\widetilde{\mathcal{H}}_{\text{CCZ}}=PH_{\text{CCZ}}P=-%
\sum_{j=2}^{2L-2}K_{j}^{\text{CCZ}}.
\end{equation}%
The energy spectrum for an open chain is shown in Fig.\ref{FigCCZEne}(b).
There is no operator for the qubits with $j=1,2,2L-1,2L$ in the Hamiltonian.
Then, the ground state is given by%
\begin{equation}
\left\vert \psi \right\rangle =\left\vert s_{1}\right\rangle \otimes
\left\vert s_{2}\right\rangle \otimes \left(
\bigotimes_{j=3}^{2L-2}\left\vert +\right\rangle \right) \otimes \left\vert
s_{2L-1}\right\rangle \otimes \left\vert s_{2L}\right\rangle .
\end{equation}

There are 16-fold degenerate states for an open chain because there are only 
$2L-2$ stabilizers. It indicates that there are four free spins at edges,
where two free spins at one edge. They used as a symmetry-protected
topological logical $2$-qubit system. They are explicitly given by 
\begin{align}
\text{X}_{\text{left}}^{1}\equiv & \text{CCZ}_{1,2,3}\text{X}_{1}\text{CCZ}%
_{1,2,3}=\text{X}_{1}\text{CZ}_{2,3},\quad \\
\text{Y}_{\text{left}}^{1}\equiv & \text{CCZ}_{1,2,3}\text{Y}_{1}\text{CCZ}%
_{1,2,3}=\text{Y}_{1}\text{CZ}_{2,3},\quad \\
\text{Z}_{\text{left}}^{1}\equiv & \text{CCZ}_{1,2,3}\text{Z}_{1}\text{CCZ}%
_{1,2,3}=\text{Z}_{1},
\end{align}%
and%
\begin{align}
\text{X}_{\text{left}}^{2}\equiv & \text{CCZ}_{2,3,4}\text{CCZ}_{1,2,3}\text{%
X}_{2}\text{CCZ}_{1,2,3}\text{CCZ}_{2,3,4}  \notag \\
=& \text{X}_{2}\text{CZ}_{2,3}\text{CZ}_{3,4},\quad \\
\text{Y}_{\text{left}}^{2}\equiv & \text{CCZ}_{2,3,4}\text{CCZ}_{1,2,3}\text{%
Y}_{2}\text{CCZ}_{1,2,3}\text{CCZ}_{2,3,4}  \notag \\
=& \text{Y}_{2}\text{CZ}_{2,3}\text{CZ}_{3,4},\quad \\
\text{Z}_{\text{left}}^{2}\equiv & \text{CCZ}_{2,3,4}\text{CCZ}_{1,2,3}\text{%
Z}_{2}\text{CCZ}_{1,2,3}\text{CCZ}_{2,3,4}  \notag \\
=& \text{Z}_{2}
\end{align}%
for the left edge, while 
\begin{align}
\text{X}_{\text{right}}^{1}\equiv & \text{CCZ}_{2L-2,2L-1,2L}\text{X}_{2L}%
\text{CCZ}_{2L-2,2L-1,2L}  \notag \\
=& \text{X}_{2L}\text{CZ}_{2L-1},\quad \\
\text{Y}_{\text{right}}^{1}\equiv & \text{CCZ}_{2L-2,2L-1,2L}\text{Y}_{2L}%
\text{CCZ}_{2L-2,2L-1,2L}  \notag \\
=& \text{Y}_{2L}\text{CZ}_{2L-1},\quad \\
\text{Z}_{\text{right}}^{1}\equiv & \text{CCZ}_{2L-2,2L-1,2L}\text{Z}_{2L}%
\text{CCZ}_{2L-2,2L-1,2L}  \notag \\
=& \text{Z}_{2L},
\end{align}%
and%
\begin{align}
\text{X}_{\text{right}}^{2}\equiv & \text{CCZ}_{2L-2,2L-1,2L}\text{CCZ}%
_{2L-3,2L-2,2L-1}  \notag \\
& \times \text{X}_{2L-1}\text{CCZ}_{2L-3,2L-2,2L-1}\text{CCZ}_{2L-2,2L-1,2L}
\notag \\
=& \text{X}_{2L-1}\text{CZ}_{2L,2L-2}\text{CZ}_{2L-1,2L-2},\quad \\
\text{Y}_{\text{right}}^{2}\equiv & \text{CCZ}_{2L-2,2L-1,2L}\text{CCZ}%
_{2L-3,2L-2,2L-1}  \notag \\
& \times \text{Y}_{2L-1}\text{CCZ}_{2L-3,2L-2,2L-1}\text{CCZ}_{2L-2,2L-1,2L}
\notag \\
=& \text{Y}_{2L-1}\text{CZ}_{2L,2L-2}\text{CZ}_{2L-1,2L-2},\quad \\
\text{Z}_{\text{right}}^{2}\equiv & \text{CCZ}_{2L-2,2L-1,2L}\text{CCZ}%
_{2L-3,2L-2,2L-1}  \notag \\
& \times \text{Z}_{2L-1}\text{CCZ}_{2L-3,2L-2,2L-1}\text{CCZ}_{2L-2,2L-1,2L}
\notag \\
=& \text{Z}_{2L-1}
\end{align}%
for the right edge.

\subsection{Topological phase transition}

As in the case of the ZXZ model, we consider a Hamiltonian interpolating $%
\mathcal{H}_{\text{CCZ}}$ and $\mathcal{H}_{\text{X}}$,%
\begin{equation}
H\left( \alpha \right) =\left( 1-\alpha \right) \mathcal{H}_{\text{CCZ}%
}+\alpha \mathcal{H}_{\text{X}}.
\end{equation}%
It has a duality%
\begin{equation}
\text{CCZ}H\left( \alpha \right) \text{CCZ}=H\left( 1-\alpha \right) ,
\end{equation}%
where it is self dual at $\alpha =1/2$. There emerges an enhanced symmetry $%
\mathbb{Z}_{2}^{\text{CCZ}}$ symmetry at $\alpha =1/2$. On the other hand,
the original $Z_{2}^{\text{even}}\times Z_{2}^{\text{odd}}$ symmetry is
broken for $\alpha >0$. The energy spectrum for a closed chain is shown as a
function of $\alpha $ in Fig.\ref{FigTra}(a2). The energy spectrum is
symmetric at $\alpha =1/2$, showing the duality relation. As opposed to the
ZXZ model, the gap between the ground state and the first-excited state does
not close at $\alpha =1/2$. The energy spectrum becomes asymmetric for an
open chain as shown in Fig.\ref{FigTra}(b2) because $\mathbb{Z}_{2}^{\text{CZ%
}}$ symmetry is broken at the edges, and the ground state degeneracies are
different between $\mathcal{H}_{\text{ZXZ}}$ and $\mathcal{H}_{\text{X}}$.
The 16-fold degenerate edge states split to 2-fold degenerate edge states
for $\alpha >0$. It is because there are two spins at one edge, which
interfere between them by $\mathcal{H}_{\text{X}}$.

The string order parameter is shown in Fig.\ref{FigTra}(c2) as a function of 
$\alpha $. It monotonically decreases as the increase of $\alpha $. However,
there is no jump at $\alpha =1/2$. These facts indicate that a topological
phase transition occurs at $\alpha =0$, which is contrasted to the ZXZ model
as in Fig.\ref{FigTra}. 

\subsection{Measurement-based quantum computation}

MBQC based on the CCZ gates is known to be universal\cite{Gach,Takeuchi}. We
explicitly construct a universal set of quantum gates; the one-qubit
universal quantum gate and the CZ gate. The former is implemented in the
standard MBQC based on the CZ gate. On the other hand, it is impossible to
implement the CZ gate in a pure one-dimensional system based on the standard
MBQC.

\subsubsection{One-qubit universal quantum gate}

We consider a case where we act one-qubit quantum gate to the second qubit.
We set three-qubit state as the initial state%
\begin{equation}
\lvert 1\rangle _{1}\otimes \lvert \psi \rangle _{2}\otimes \lvert +\rangle
_{3}.
\end{equation}%
Here, the first qubit is set to be 1 and the third qubit to be the state $%
\lvert +\rangle _{3}$. We take the second qubit as the input%
\begin{equation}
\lvert \psi \rangle _{2}=\sum_{x_{2}}\psi _{x_{2}}\,\lvert x_{2}\rangle _{2},
\end{equation}%
where $\lvert \psi \rangle _{2}$ is the state for the second qubit, $%
x_{2}=0,1$ represents the qubit, $\psi _{x_{2}}$ is the complex number, $%
\lvert x_{2}\rangle _{2}$ is the eigenstate for the second qubit. Then, the
CCZ gate is reduced to the standard CZ gate, 
\begin{align}
\lvert \Psi \rangle _{123}=& \mathrm{CCZ}_{123}\bigl(\lvert 1\rangle
_{1}\otimes \lvert \psi \rangle _{2}\otimes \lvert +\rangle _{3}\bigr) 
\notag \\
=& \lvert 1\rangle _{1}\otimes \mathrm{CZ}_{23}\bigl(\lvert \psi \rangle
_{2}\otimes \lvert +\rangle _{3}\bigr)  \notag \\
=& \lvert 1\rangle _{1}\otimes \sum_{x_{2}}\psi _{x_{2}}\,\lvert
x_{2}\rangle _{2}\frac{\lvert 0\rangle _{3}+(-1)^{x_{2}}\lvert 1\rangle _{3}%
}{\sqrt{2}}.
\end{align}%
Now, we can use the standard protocol to make a universal quantum gate. We
measure the second qubit with the angle $\theta $, where the basis
eigenfunctions are given by%
\begin{equation}
\lvert \alpha _{s}(\theta )\rangle =\frac{1}{\sqrt{2}}\bigl(\lvert 0\rangle
+(-1)^{s}e^{i\theta }\lvert 1\rangle \bigr),\quad s\in \{0,1\}.
\end{equation}%
The resultant state after the measurement is given by taking an inner
produce with respect to the second qubit as

\begin{align}
& \lvert \psi _{\text{out}}^{(s)}\rangle _{3}  \notag \\
=& _{2}\langle \alpha _{s}(\theta )\lvert \Psi \rangle _{123}  \notag \\
=& \lvert 1\rangle _{1}\otimes {}_{2}\langle \alpha _{s}(\theta )\rvert 
\notag \\
& \times \left[ \psi _{0}\,\lvert 0\rangle _{2}\frac{\lvert 0\rangle
_{3}+\lvert 1\rangle _{3}}{\sqrt{2}}+\psi _{1}\,\lvert 1\rangle _{2}\frac{%
\lvert 0\rangle _{3}-\lvert 1\rangle _{3}}{\sqrt{2}}\right]  \notag \\
=& \lvert 1\rangle _{1}\otimes \Bigl[\psi _{0}\,\frac{\lvert 0\rangle
_{3}+\lvert 1\rangle _{3}}{\sqrt{2}}+\frac{(-1)^{s}e^{-i\theta }}{\sqrt{2}}%
\psi _{1}\,\frac{\lvert 0\rangle _{3}-\lvert 1\rangle _{3}}{\sqrt{2}}\Bigr] 
\notag \\
=& \lvert 1\rangle _{1}\otimes \frac{1}{\sqrt{2}}\Bigl[\left\{ \psi _{0}\,+%
\frac{(-1)^{s}e^{-i\theta }}{\sqrt{2}}\psi _{1}\right\} \lvert 0\rangle _{3}
\notag \\
& +\left\{ \psi _{0}\,-\frac{(-1)^{s}e^{-i\theta }}{\sqrt{2}}\psi
_{1}\right\} \lvert 1\rangle _{3}\Bigr],
\end{align}%
If $s=0$, it is generated by the unitary gate%
\begin{equation}
\lvert \psi _{\text{out}}^{(0)}\rangle _{3}=U^{\left( 0\right) }\left(
\theta \right) \left\vert \psi \right\rangle _{2}
\end{equation}%
with%
\begin{equation}
U^{\left( 0\right) }\left( \theta \right) \equiv e^{-i\theta /2}\text{H}%
e^{i\theta Z/2}.  \label{U0}
\end{equation}%
First, by setting $\theta =0$, we obtain the Hadamard gate as%
\begin{equation}
U^{\left( 0\right) }\left( 0\right) =\text{H}.
\end{equation}%
Then, by multiplying the Hadamard gate to $U^{\left( 0\right) }\left( \theta
\right) $ again, we obtain the phase-shift gate with arbitrary angle $\theta 
$ as%
\begin{equation}
U^{\left( 0\right) }\left( 0\right) U^{\left( 0\right) }\left( \theta
\right) =e^{-i\theta /2}e^{i\theta Z/2}.
\end{equation}%
The Hadamard gate H and the phase-shift gate $e^{i\theta Z/2}$ with
arbitrary angle $\theta $ constitute the one-qubit unitary quantum gate.

On the other hand, if $s=1$, we have%
\begin{equation}
\lvert \psi _{\text{out}}^{(1)}\rangle _{3}=\frac{1}{2}\psi _{0}\Bigl[%
1-e^{-i\theta }\Bigr]\lvert 0\rangle _{2}+\frac{1}{2}\psi _{1}\Bigl[%
1+e^{-i\theta }\Bigr]\lvert 1\rangle _{2},
\end{equation}%
which is generated by the unitary gate%
\begin{equation}
\lvert \psi _{\text{out}}^{(0)}\rangle _{3}=U^{\left( 1\right) }\left\vert
\psi \right\rangle _{2}
\end{equation}%
with%
\begin{equation}
U^{\left( 1\right) }\left( \theta \right) \equiv e^{-i\theta /2}\text{XH}%
e^{i\theta Z/2}.  \label{U1}
\end{equation}%
There is a byproduct operator X in Eq.(\ref{U1}) in comparison to Eq.(\ref%
{U0}). This X operator is deleted by using the feed-forward control, where
we set the angle $\theta _{2}\rightarrow -\theta _{2}$ for the next
measurement because of the relations 
\begin{align}
\text{X}\lvert \alpha _{0}(\theta )\rangle =& \lvert \alpha _{1}(\theta
)\rangle , \\
\text{X}\lvert \alpha _{1}(\theta )\rangle =& \lvert \alpha _{0}(\theta
)\rangle .
\end{align}

Next, we consider a case where we act one-qubit quantum gate to the first
qubit. In this case, it is enough to set three-qubit state as the initial
state%
\begin{equation}
\lvert \psi \rangle _{1}\otimes \lvert 1\rangle _{2}\otimes \lvert +\rangle
_{3}.
\end{equation}

\subsubsection{CS gate}

We prepare an initial two-qubit states at the first and second qubits as%
\begin{equation}
\lvert \psi \rangle _{12}=\sum_{x_{1},x_{2}\in \{0,1\}}\psi
_{x_{1}x_{2}}\,\lvert x_{1}x_{2}\rangle
\end{equation}%
and the $+$ state at the third qubit as%
\begin{equation}
\lvert +\rangle _{3}=\frac{1}{\sqrt{2}}(\lvert 0\rangle +\lvert 1\rangle ).
\end{equation}%
Then, the CCZ cluster state is obtained by applying the CCZ gate as%
\begin{align}
\lvert \Psi \rangle _{123}=& \mathrm{CCZ}_{123}\bigl(\lvert \psi \rangle
_{12}\otimes \lvert +\rangle _{3}\bigr)  \notag \\
=& \sum_{x_{1},x_{2}}\psi _{x_{1}x_{2}}\,\lvert x_{1}x_{2}\rangle _{12}\left[
\lvert 0\rangle _{3}+(-1)^{x_{1}x_{2}}\lvert 1\rangle _{3}\right] .
\end{align}%
We measure the third qubit with the angle $\theta $%
\begin{equation}
\lvert \alpha _{s}(\theta )\rangle =\frac{1}{\sqrt{2}}\bigl(\lvert 0\rangle
+(-1)^{s}e^{i\theta }\lvert 1\rangle \bigr),\quad s\in \{0,1\}.
\end{equation}%
The resultant state after the measurement is given by taking an inner
produce with respect to the third qubit as

\begin{align}
& \lvert \psi _{\text{out}}^{(s)}\rangle _{12}  \notag \\
=& _{3}\langle \alpha _{s}(\theta )\lvert \Psi \rangle _{123}  \notag \\
=& \sum_{x_{1},x_{2}}\psi _{x_{1}x_{2}}\,\lvert x_{1}x_{2}\rangle \frac{1}{%
\sqrt{2}}\Bigl(\langle \alpha _{s}(\theta )|0\rangle
+(-1)^{x_{1}x_{2}}\langle \alpha _{s}(\theta )|1\rangle \Bigr)  \notag \\
=& \left( \sum_{x_{1},x_{2}}\psi _{x_{1}x_{2}}\,\lvert x_{1}x_{2}\rangle
\right) \frac{1}{\sqrt{2}}\left( \frac{1}{\sqrt{2}}+(-1)^{x_{1}x_{2}}\frac{%
(-1)^{s}e^{-i\theta }}{\sqrt{2}}\right)  \notag \\
=& \frac{1}{2}\sum_{x_{1},x_{2}}\psi _{x_{1}x_{2}}\Bigl[%
1+(-1)^{x_{1}x_{2}+s}e^{-i\theta }\Bigr]\lvert x_{1}x_{2}\rangle ,
\end{align}%
where we have used the relations%
\begin{equation}
\langle \alpha _{s}(\theta )|0\rangle =\frac{1}{\sqrt{2}},\quad \langle
\alpha _{s}(\theta )|1\rangle =\frac{(-1)^{s}e^{-i\theta }}{\sqrt{2}}.
\end{equation}%
The corresponding unitary gate $U_{s}(\theta )$ is defined by%
\begin{equation}
\lvert \psi _{\text{out}}^{(s)}\rangle _{12}=\frac{1}{2}\,U_{s}(\theta
)\,\lvert \psi \rangle _{12},
\end{equation}%
which is explicitly calculated as%
\begin{equation}
U_{s}(\theta )=\sum_{x_{1},x_{2}}\Bigl[1+(-1)^{x_{1}x_{2}+s}e^{-i\theta }%
\Bigr]\lvert x_{1}x_{2}\rangle \langle x_{1}x_{2}\rvert .
\end{equation}%
By introducing%
\begin{align}
\lambda _{0,s}(\theta )=& 1+(-1)^{s}e^{-i\theta }, \\
\lambda _{1,s}(\theta )=& 1-(-1)^{s}e^{-i\theta },
\end{align}%
it is explicitly given by the diagonal gate as

\begin{align}
U_{s}(\theta )=& \mathrm{diag}\bigl(\lambda _{0,s},\lambda _{0,s},\lambda
_{0,s},\lambda _{1,s}\bigr)  \notag \\
=& \lambda _{0,s}\mathrm{diag}\bigl(1,1,1,\lambda _{s}(\theta )\bigr)
\end{align}%
with%
\begin{equation}
\lambda _{s}(\theta )=\frac{\lambda _{1,s}(\theta )}{\lambda _{0,s}(\theta )}%
.
\end{equation}%
If we set $\theta =\pi /2$, we obtain the CS gate%
\begin{equation}
U_{s}(\theta )=\lambda _{0,s}\text{CS}
\end{equation}%
with%
\begin{equation}
\text{CS}=\text{diag.}\left( 1,1,1,si\right) .
\end{equation}%
The square of the CS gate is the CZ gate CZ$=($CS$)^{2}$. Then, we have
one-qubit universal quantum gate and the CZ gate, which enable us universal
quantum computation.

Actually, three-qubits are enough for MBQC if we also use the CZ gate. In
order to apply a one-qubit quantum gate to the first qubit, we apply the CZ
to the first qubit and the third qubit 
\begin{equation}
\text{CZ}_{13}\lvert \psi \rangle _{12}\otimes \lvert +\rangle _{3}.
\end{equation}%
Similarly, in order to apply a one-qubit quantum gate to the second qubit,
we apply the CZ to the second qubit and the third qubit 
\begin{equation}
\text{CZ}_{23}\lvert \psi \rangle _{12}\otimes \lvert +\rangle _{3}.
\end{equation}%
On the other hand, in order to apply a two-qubit quantum gate to the first
qubit and the second qubit, we apply the CCZ to the first qubit, the second
qubit and the third qubit 
\begin{equation}
\text{CCZ}_{123}\lvert \psi \rangle _{12}\otimes \lvert +\rangle _{3}.
\end{equation}%
After the measurement, the active qubits are two-qubit state $\lvert \psi
\rangle _{12}$ because one qubit is discarded by the projective measurement.
The next quantum gate operation is executed by initializing the discarded
qubit to the state $\lvert +\rangle _{3}$. Hence, any number of quantum gate
operations are executed solely by three qubits.

\section{C$^{N}$Z gate model}

We consider a hyper-cluster state generated by the C$^{N}$Z gate\cite{Rossi},%
\begin{equation}
\left\vert \psi \right\rangle =\prod_{j=1}^{2L-N}\text{C}^{N}\text{Z}%
_{\left\{ j;N\right\} }\bigotimes_{j=1}^{2L}\left\vert +\right\rangle _{j},
\label{psiCNZ}
\end{equation}%
which is of the form of Eq.(\ref{UjN}), where the C$^{N}$Z gate is defined by%
\begin{equation}
\text{C}^{N}\text{Z}_{\left\{ j;N\right\} }\equiv 1-\bigotimes_{k=0}^{N}%
\frac{1-\text{Z}_{j+k}}{2},
\end{equation}%
and $\left\{ j;N\right\} \equiv \left\{ j,j+1,\cdots ,j+N\right\} $
represents the $N$ sequential numbers starting from $j$. The generated
cluster states are called the hyper-graph states\cite{Rossi}. The action of
the C$^{N}$Z gate is%
\begin{equation}
\text{C}^{N}\text{Z}_{\left\{ j;N\right\} }\left\vert x_{0}x_{1}\cdots
x_{N}\right\rangle =\left( -1\right) ^{x_{0}x_{1}\cdots x_{N}}\left\vert
x_{0}x_{1}\cdots x_{N}\right\rangle 
\end{equation}%
with $x_{j}=0,1$ for $j=0,1,\cdots N$.

The corresponding Hamiltonian is given by%
\begin{equation}
\mathcal{H}_{\text{C}^{N}\text{Z}}=-\sum_{j}K_{j}^{\text{C}^{N}\text{Z}}
\end{equation}%
with the stabilizer%
\begin{equation}
K_{j}^{\text{C}^{N}\text{Z}}=\text{C}^{N}\text{Z}_{\left\{ j\right\} }\text{X%
}_{j}\text{C}^{N}\text{Z}_{\left\{ j\right\} },
\end{equation}%
where%
\begin{equation}
\text{C}^{N}\text{Z}_{\left\{ j\right\} }\equiv \prod\limits_{k=0}^{N}\text{%
C}^{N}\text{Z}_{\left\{ j-N+k;N\right\} }.
\end{equation}%
It is explicitly written in the form of%
\begin{align}
& K_{j}^{\text{C}^{N}\text{Z}}  \notag \\
=& \sum_{\left\{ \zeta _{j}\right\} }\alpha _{\zeta _{j-N}\cdots \zeta
_{j-1}\zeta _{j+1}\cdots \zeta _{j+N}}\left( \prod\limits_{n=1}^{N}\text{Z}%
_{j-n}^{\zeta _{j-n}}\right) \text{X}_{j}\prod\limits_{n=1}^{N}\text{Z}%
_{j+n}^{\zeta _{j+n}},
\end{align}%
where $\zeta _{j}=0,1$ and the coefficient is%
\begin{equation}
\alpha _{\zeta _{j-N}\cdots \zeta _{j-1}\zeta _{j+1}\cdots \zeta _{j+N}}=\pm 
\frac{1}{2^{N}}.
\end{equation}%
The C$^{N}$Z$_{j,j+1}$ gate is not a Clifford gate for $N\geq 2$. The
coefficients are determined as in the case of the CCZ gate by using the $N$%
-body Z$^{N}$ interactions. The generalization to the C$^{N}$P gate is
straightforward and the stabilizer has the same form with different
coefficients.

The $\mathbb{Z}_{2}^{\text{even}}\times \mathbb{Z}_{2}^{\text{odd}}$\
symmetry is written in the form of%
\begin{equation}
\eta _{\text{C}^{N}\text{Z}}^{\text{even}}=\prod\limits_{j\in \text{even}%
}\prod_{k=0}^{N}\text{C}^{N-1}\text{Z}_{\left\{ j\right\} _{k}^{N}-j}\text{X%
}_{j},
\end{equation}%
where $\left\{ j\right\} _{k}^{N}\equiv \left\{ j-N+k;N\right\} $, while $%
\left\{ j\right\} _{k}^{N}-j$ represents the difference set of $\left\{
j\right\} _{k}^{N}$ and $j$. In this derivation, we have used the relation%
\begin{align}
& \left( \prod\limits_{k=0}^{N}\text{C}^{N}\text{Z}_{\left\{ j\right\}
_{k}^{N}}\right) \text{X}_{j}\left( \prod\limits_{k=0}^{N}\text{C}^{N}\text{%
Z}_{\left\{ j\right\} _{k}^{N}}\right)   \notag \\
=& \text{X}_{j}\prod_{k=0}^{N}\text{C}^{N-1}\text{Z}_{\left\{ j\right\}
_{k}^{N}-j}.
\end{align}%
Hence, it is not in the Clifford group for $N\geq 3$ because the C$^{N}$Z
gate is not a Clifford gate for $N\geq 2$.

There is a unique ground state for a closed chain because there are $L$
stabilizers $K_{j}^{\text{C}^{N}\text{Z}}$ for $1\leq j\leq L$, where $j$ is
defined modulo $L$.

\subsection{Edge states}

We consider the $\mathbb{Z}_{2}^{\text{even}}\times \mathbb{Z}_{2}^{\text{odd%
}}$ symmetry projected Hamiltonian,%
\begin{equation}
\widetilde{H}_{\text{C}^{N}\text{Z}}=PH_{\text{C}^{N}\text{Z}%
}P=-\sum_{j=N}^{2L-N}K_{\left\{ j\right\} }^{\text{C}^{N}\text{Z}}.
\end{equation}%
We make a unitary transformation%
\begin{align}
& \text{C}^{N}\text{Z}\mathcal{H}_{\text{ZXZ}}\text{C}^{N}\text{Z}  \notag \\
=& -\sum_{j=N+1}^{2L-N-1}\left( \prod\limits_{k=0}^{N}\text{C}^{N-1}\text{Z}%
_{\left\{ j\right\} _{k}^{N}-j}\right) \text{X}_{j}\prod\limits_{k=0}^{N}%
\text{C}^{N-1}\text{Z}_{\left\{ j\right\} _{k}^{N}-j}.
\end{align}%
The ground state is explicitly given by%
\begin{equation}
\left\vert \psi \right\rangle =\left( \bigotimes_{j=1}^{N}\left\vert
s_{j}\right\rangle \right) \otimes \left(
\bigotimes_{j=N+1}^{2L-N-1}\left\vert +\right\rangle \right) \otimes \left(
\bigotimes_{j=2L-N}^{2L}\left\vert s_{j}\right\rangle \right) .
\end{equation}%
There are $2^{2N}$-fold degenerate states for an open chain because there
are only $L-2N$ stabilizers $K_{j}^{\text{C}^{N}\text{Z}}$ for $N+1\leq
j\leq L-N$. It shows that there are $N$ free spins at each edge. They are
used as a symmetry-protected topological $N$-qubit system protected by the $%
\mathbb{Z}_{2}^{\text{even}}\times \mathbb{Z}_{2}^{\text{odd}}$ symmetry.

The logical qubits are explicitly obtained as%
\begin{align}
\text{X}_{\text{left}}^{1}=& \text{C}^{N}\text{Z}_{1,\cdots ,N}\text{X}_{1}%
\text{C}^{N}\text{Z}_{1,\cdots ,N}=\text{X}_{1}\text{C}^{N-1}\text{Z}%
_{2,\cdots N}, \\
\text{Y}_{\text{left}}^{1}=& \text{C}^{N}\text{Z}_{1,\cdots ,N}\text{Y}_{1}%
\text{C}^{N}\text{Z}_{1,\cdots ,N}=\text{Y}_{1}\text{C}^{N-1}\text{Z}%
_{2,\cdots N},
\end{align}%
\begin{align}
\text{X}_{\text{left}}^{2}=& \text{C}^{N}\text{Z}_{2,\cdots ,N+1}\text{C}^{N}%
\text{Z}_{1,\cdots ,N}\text{X}_{2}\text{C}^{N}\text{Z}_{2,\cdots ,N+1} 
\notag \\
=& \text{X}_{2}\text{C}^{N-1}\text{Z}_{2,\cdots ,N}\text{C}^{N-1}\text{Z}%
_{1,\cdots ,N-1}, \\
\text{Y}_{\text{left}}^{2}=& \text{C}^{N}\text{Z}_{2,\cdots ,N+1}\text{C}^{N}%
\text{Z}_{1,\cdots ,N}\text{Y}_{2}\text{C}^{N}\text{Z}_{2,\cdots ,N+1} 
\notag \\
=& \text{Y}_{1}\text{C}^{N-1}\text{Z}_{2,\cdots ,N}\text{C}^{N-1}\text{Z}%
_{1,\cdots ,N-1},
\end{align}%
\begin{equation*}
\vdots
\end{equation*}

\begin{align}
\text{X}_{\text{left}}^{n}=& \prod\limits_{j}\text{C}^{N-1}\text{Z}%
_{\left\{ j;N\right\} -n}\text{X}_{n}=\text{X}_{n}\prod\limits_{j}\text{C}%
^{N-1}\text{Z}_{\left\{ j;N\right\} -n}, \\
\text{Y}_{\text{left}}^{n}=& \prod\limits_{j}\text{C}^{N-1}\text{Z}%
_{\left\{ j;N\right\} -n}\text{X}_{n}=\text{Y}_{n}\prod\limits_{j}\text{C}%
^{N-1}\text{Z}_{\left\{ j;N\right\} -n},
\end{align}%
and 
\begin{equation}
\text{Z}_{\text{left}}^{j}=\text{Z}_{j}.
\end{equation}

\subsection{Measurement-based quantum computation and C$^{N-1}$S gate}

We prepare an initial $N$-qubit states 
\begin{equation}
\lvert \psi \rangle _{1,2,\cdots ,N-1,N}=\sum_{x_{1},x_{2}\in \{0,1\}}\psi
_{x_{1}x_{2}\cdots x_{N}}\,\lvert x_{1}x_{2}\cdots x_{N}\rangle
\end{equation}%
and the $+$ state at the $\left( N+1\right) $-th qubit as 
\begin{equation}
\lvert +\rangle _{N}=\frac{\lvert 0\rangle _{N}+\lvert 1\rangle _{N}}{\sqrt{2%
}}.
\end{equation}%
Then, the CCZ cluster state is obtained by applying the CCZ gate as%
\begin{align}
& \lvert \Psi \rangle _{1,2,\cdots ,N+1}  \notag \\
=& \mathrm{C}^{N-1}\mathrm{Z}_{1,2,\cdots ,N+1}\bigl(\lvert \psi \rangle
_{12,\cdots ,N}\otimes \lvert +\rangle _{N+1}\bigr)  \notag \\
=& \sum_{x_{1},x_{2},\cdots ,x_{N}}\psi _{x_{1}x_{2}\cdots x_{N}}\,\lvert
x_{1}x_{2}\cdots x_{N}\rangle _{12\cdots ,N}  \notag \\
& \times \left[ \lvert 0\rangle _{N+1}+(-1)^{x_{1}x_{2}\cdots x_{N}}\lvert
1\rangle _{N+1}\right] .
\end{align}%
We perform a measurement of the $\left( N+1\right) $-th qubit in the basis%
\begin{equation}
\lvert \alpha _{s}(\theta )\rangle =\frac{1}{\sqrt{2}}\bigl(\lvert 0\rangle
+(-1)^{s}e^{i\theta }\lvert 1\rangle \bigr),\quad s\in \{0,1\}.
\end{equation}%
The resultant state after the measurement is given by taking an inner
produce with respect to the ($N+1$)-th qubit as

\begin{align}
& \lvert \psi _{\text{out}}^{(s)}\rangle _{1,2,\cdots ,N}  \notag \\
=& {}_{N+1}\langle \alpha _{s}(\theta )\lvert \Psi \rangle _{1,2\cdots ,N} 
\notag \\
=& \sum_{x_{1},x_{2},\cdots ,x_{N}}\psi _{x_{1}x_{2}\cdots x_{N}}\,\lvert
x_{1}x_{2}\cdots x_{N}\rangle  \notag \\
& \times \frac{1}{\sqrt{2}}\Bigl(\langle \alpha _{s}(\theta )|0\rangle
+(-1)^{x_{1}x_{2}\cdots x_{N}}\langle \alpha _{s}(\theta )|1\rangle \Bigr) 
\notag \\
=& \left( \sum_{x_{1},x_{2},\cdots ,x_{N}}\psi _{x_{1}x_{2}\cdots
x_{N}}\,\lvert x_{1}x_{2}\cdots x_{N}\rangle \right)  \notag \\
& \times \frac{1}{\sqrt{2}}\left( \frac{1}{\sqrt{2}}+(-1)^{x_{1}x_{2}\cdots
x_{N}}\frac{(-1)^{s}e^{-i\theta }}{\sqrt{2}}\right)  \notag \\
=& \frac{1}{2}\sum_{x_{1},x_{2},\cdots ,x_{N}}\psi _{x_{1}x_{2}\cdots x_{N}}%
\Bigl[1+(-1)^{x_{1}x_{2}+s}e^{-i\theta }\Bigr]\lvert x_{1}x_{2}\cdots
x_{N}\rangle ,
\end{align}%
where we have used the relations%
\begin{equation}
\langle \alpha _{s}(\theta )|0\rangle =\frac{1}{\sqrt{2}},\quad \langle
\alpha _{s}(\theta )|1\rangle =\frac{(-1)^{s}e^{-i\theta }}{\sqrt{2}}.
\end{equation}%
The corresponding unitary gate $U_{s}(\theta )$ is defined by%
\begin{equation}
\lvert \psi _{\text{out}}^{(s)}\rangle _{1,2,\cdots ,N}=\frac{1}{2}%
\,U_{s}(\theta )\,\lvert \psi \rangle _{1,2,\cdots ,N},
\end{equation}%
which is explicitly calculated as%
\begin{equation}
U_{s}(\theta )=\sum_{x_{1},x_{2},\cdots ,x_{N}}\Bigl[1+(-1)^{x_{1}x_{2}%
\cdots x_{N}+s}e^{-i\theta }\Bigr]\lvert x_{1}x_{2}\rangle \langle
x_{1}x_{2}\rvert .
\end{equation}%
By introducing%
\begin{align}
\lambda _{0,s}(\theta )=& 1+(-1)^{s}e^{-i\theta }, \\
\lambda _{1,s}(\theta )=& 1-(-1)^{s}e^{-i\theta },
\end{align}%
it is explicitly given by the diagonal gate as

\begin{align}
U_{s}(\theta )=& \mathrm{diag}\bigl(\lambda _{0,s},\lambda _{0,s},\lambda
_{0,s},\lambda _{1,s}\bigr)  \notag \\
=& \lambda _{0,s}\mathrm{diag}\bigl(1,1,1,\lambda _{s}(\theta )\bigr)
\end{align}%
with%
\begin{equation}
\lambda _{s}(\theta )=\frac{\lambda _{1,s}(\theta )}{\lambda _{0,s}(\theta )}%
.
\end{equation}%
If we set $\theta =\pi /2$, we obtain the CS gate%
\begin{equation}
U_{s}(\theta )=\lambda _{0,s}\text{CS}
\end{equation}%
with%
\begin{equation}
\text{C}^{N-1}\text{S}=\text{diag.}\left( 1,1,1,\cdots ,si\right) .
\end{equation}%
The square of the C$^{N-1}$S gate is the C$^{N-1}$Z gate, C$^{N-1}$Z$=\left( 
\text{C}^{N-1}\text{S}\right) ^{2}$. In general, it is hard to directly
implement C$^{N-1}$Z gate in quantum computation.

As in the case of the CCZ hyper-cluster state, $N$ qubits are enough for
MBQC, where any number of quantum gates are executed by initialization and
measurement of $N$-th qubit.

\section{Discussions}

We have shown that symmetry-protected $N$-fold degenerate edge states emerge in hyper-cluster models. These edge degrees of freedom can be used as an $N$ qubits in MBQC.
We have explicitly constructed the CZ gate within the CCZ cluster state, thereby 
realizing universal quantum computation. Furthermore, we have demonstrated that the C$%
^{N-1}Z$ gate can be directly implemented using the C$^{N}Z$ cluster state.

We discuss experimental realizations. The CP gate is constructed by using
the Ising gate as shown in Appendix \ref{ApIsing}. Especially, the
cross-resonance gate\cite{CorcoS} is used in transmon-type superconducting
qubits\cite{Chow,Corco} and the M\o lmer-S\o rensen gate\cite{MSS} is used
in qubits made of ion trap\cite{Kaler}. It is generated in photonic qubits%
\cite{Xu} and photonic systems\cite{Krast}. The CCZ gate is generated by
entangled photons\cite{Wang}, superconducting qubits\cite{LiuCCZ} and
Rydberg Ions\cite{Bols}. The CCP gate is generated in superconducting qubits%
\cite{Glazer}. The C$^{N}$Z gate is generated in Rydberg Ions\cite{Pele,Xue}.

\section*{Acknowledgements}

The author is grateful to R. Kobayashi, H. Watanabe, H. Yagi and H. Oshima
for helpful discussions on the subject. This work is supported by CREST, JST
(Grants No. JPMJCR20T2) and Grants-in-Aid for Scientific Research from MEXT
KAKENHI (Grant No. 23H00171).

\section*{Appendices}

\appendix

\section{ZXZ model}

\label{ApZXZ}

We summerize the properties of the ZXZ model for the closed and open chains,
symmetry properties and the string-order parameter.

\subsection{Closed chain}

For a closed chain, we consider the Hamiltonian (\ref{HZXZOpen})$\ $given by%
\begin{equation}
\mathcal{H}_{\text{ZXZ}}=-\sum_{j=1}^{2L}K_{j}^{\text{ZXZ}}  \label{HZXZK}
\end{equation}%
by identifying the qubit $j=2L+1$ and the qubit $j=1$, where the stabilizers
are given by%
\begin{equation}
K_{j}^{\text{ZXZ}}=\text{Z}_{j-1}\text{X}_{j}\text{Z}_{j+1}  \label{KZXZ}
\end{equation}%
for $2\leq j\leq 2L-1$ and 
\begin{align}
K_{1}^{\text{ZXZ}}=& \text{Z}_{2L}\text{X}_{1}\text{Z}_{2}, \\
K_{2L}^{\text{ZXZ}}=& \text{Z}_{2L-1}\text{X}_{2L}\text{Z}_{1}.
\end{align}%
The eigenspectrum for a closed chain is shown in Fig.\ref{FigZXZEne}(a),
where there is a single gapped ground state as indicated by a closed red
circle.

\subsection{Non-invertible symmetry}

In addition to the $\mathbb{Z}_{2}^{\text{even}}\times \mathbb{Z}_{2}^{\text{%
odd}}$ symmetry, the Hamiltonian (\ref{HZXZK}) also satisfies the
non-invertible symmetry\cite{Seiberg,Seiberg2,Mana},%
\begin{equation}
\mathsf{D}_{\text{ZXZ}}\mathcal{H}_{\text{ZXZ}}=\mathcal{H}_{\text{ZXZ}}%
\mathsf{D}_{\text{ZXZ}},
\end{equation}%
whose generator is given by%
\begin{equation}
\mathsf{D}_{\text{ZXZ}}=\mathsf{TD}_{\text{ZXZ}}^{\text{even}}\mathsf{D}_{%
\text{ZXZ}}^{\text{odd}}.  \label{DZXZ}
\end{equation}%
Here, the translation symmetry operator $\mathsf{T}$ satisfies 
\begin{equation}
\mathsf{T}\text{X}_{j}\mathsf{T}^{-1}=\text{X}_{j+1},\qquad \mathsf{T}\text{Z%
}_{j}\mathsf{T}^{-1}=\text{Z}_{j+1},
\end{equation}%
and the non-invertible symmetry operators are given by 
\begin{align}
\mathsf{D}_{\text{ZXZ}}^{\text{even}}=& e^{\frac{\pi iN}{8}}P_{\text{ZXZ}%
}e^{-\frac{i\pi }{4}\text{X}_{2L}}\mathsf{D}_{2L-2}^{\text{ZXZ}}\cdots 
\mathsf{D}_{4}^{\text{ZXZ}}\mathsf{D}_{2}^{\text{ZXZ}}, \\
\mathsf{D}_{\text{ZXZ}}^{\text{odd}}=& e^{\frac{\pi iN}{8}}P_{\text{ZXZ}}e^{-%
\frac{i\pi }{4}\text{X}_{2L-1}}\mathsf{D}_{2L-3}^{\text{ZXZ}}\cdots \mathsf{D%
}_{3}^{\text{ZXZ}}\mathsf{D}_{1}^{\text{ZXZ}}
\end{align}%
with%
\begin{equation}
\mathsf{D}_{j}^{\text{ZXZ}}\equiv e^{-\frac{i\pi }{4}\text{Z}_{j}\text{Z}%
_{j+1}}e^{-\frac{i\pi }{4}\text{X}_{j}},
\end{equation}%
and the projection operator is given by%
\begin{equation}
P_{\text{ZXZ}}\equiv \frac{1+\eta _{\text{ZXZ}}}{2}
\end{equation}%
with%
\begin{equation}
\eta _{\text{ZXZ}}\equiv \eta _{\text{ZXZ}}^{\text{even}}\eta _{\text{ZXZ}}^{%
\text{odd}}.  \label{etaZXZ2}
\end{equation}

It is noted that $\mathsf{D}_{\text{ZXZ}}^{\text{even}}$ and $\mathsf{D}_{%
\text{ZXZ}}^{\text{odd}}$ do not have inverses due to the projection
operator $P\equiv \left( 1+\eta _{\text{ZXZ}}\right) /2$ satisfying $P^{2}=P$%
, which restricts the eingenstates to preserve the $\mathbb{Z}_{2}^{\text{%
even}}\times \mathbb{Z}_{2}^{\text{odd}}$ symmetry. Hence, $\mathsf{D}_{%
\text{ZXZ}}$ is called non-invertible symmetry. The non-invertible symmetry $%
\mathsf{D}_{\text{ZXZ}}$ maps X$_{j}$ and Z$_{j-1}$Z$_{j+1}$ as%
\begin{align}
\mathsf{D}_{\text{ZXZ}}\text{X}_{j}=& \text{Z}_{j-1}\text{Z}_{j+1}\mathsf{D}%
_{\text{ZXZ}},Ex \\
\mathsf{D}_{\text{ZXZ}}\text{Z}_{j-1}\text{Z}_{j+1}=& \text{X}_{j}\mathsf{D}%
_{\text{ZXZ}},
\end{align}%
which are abbreviated as%
\begin{equation}
\text{X}_{j}\rightsquigarrow \text{Z}_{j-1}\text{Z}_{j+1},\qquad \text{Z}%
_{j-1}\text{Z}_{j+1}\rightsquigarrow \text{X}_{j},
\end{equation}%
implying that the ZXZ model is self dual under the non-invertible symmetry.

\subsection{Kennedy-Tasaki transformation}

The Kennedy-Tasaki (KT) transformation\cite{KT1,KT2,Oshikawa} maps the SPT
model to the Ising model with the next-nearest neighbor interaction,%
\begin{equation}
H_{\text{Ising}}\equiv \sum_{j}\text{Z}_{j-1}\text{Z}_{j+1},  \label{HIsing}
\end{equation}%
showing the emergence of a spontaneous symmetry broken phase. The KT
transformation for the ZXZ model is defined by\cite{Mana,Seif,LiOshi}%
\begin{equation}
\mathsf{KT}\equiv \left( \mathsf{D}_{\text{ZXZ}}^{\text{even}}\mathsf{D}_{%
\text{ZXZ}}^{\text{odd}}\right) ^{\dagger }\left( \text{CZ}\right) \left( 
\mathsf{D}_{\text{ZXZ}}^{\text{even}}\mathsf{D}_{\text{ZXZ}}^{\text{odd}%
}\right) ^{\dagger },  \label{KTZXZ}
\end{equation}%
which transforms%
\begin{align}
\left( \mathsf{KT}\right) \text{X}_{j}=& \text{X}_{j}\left( \mathsf{KT}%
\right) , \\
\left( \mathsf{KT}\right) \text{Z}_{j-1}\text{X}_{j}\text{Z}_{j+1}=& \text{Z}%
_{j-1}\text{Z}_{j+1}\left( \mathsf{KT}\right) .
\end{align}%
They are abbreviated as%
\begin{align}
\text{X}_{j}& \leadsto \text{X}_{j}, \\
\text{Z}_{j-1}\text{X}_{j}\text{Z}_{j+1}& \leadsto \text{Z}_{j-1}\text{Z}%
_{j+1}.
\end{align}%
Hence, the ZXZ model (\ref{HZXZK}) is transformed to the Ising model (\ref%
{HIsing}), 
\begin{equation}
H_{\text{ZXZ}}\leadsto H_{\text{Ising}},
\end{equation}%
by the KT\ transformation.

The Ising model does not have a topological phase although the ZXZ model
has. This is because the topological property is not conserved by the KT
transformation, which is a nonlocal transformation.

\subsection{String-order parameter}

It is possible to define a string-order parameter\cite{KT2} in the ZXZ model
(\ref{HZXZK}) by%
\begin{align}
\mathcal{O}\left( i,j\right) \equiv &\left\langle g\right\vert
\prod\limits_{k=i}^{j}K_{k}^{\text{ZXZ}}\left\vert g\right\rangle  \notag \\
=&\left( -1\right) ^{i-j-1}\left\langle g\right\vert \text{Z}_{i}\text{Y}%
_{i+1}\left( \prod\limits_{k=i+2}^{j-2}\text{X}_{k}\right) \text{Y}_{j-1}%
\text{Z}_{j}\left\vert g\right\rangle ,
\end{align}%
where $\left\vert g\right\rangle $ is the ground state of a given
Hamiltonian and we have used Eq.(\ref{KZXZ}). It is 1 for the SPT state $%
\left\vert g\right\rangle =\left\vert \psi ^{\text{ZXZ}}\right\rangle $,
which is the ground state of $\mathcal{H}_{\text{ZXZ}}$ due to the
stabilizer condition Eq.(\ref{stabilizerg}), while it is zero for the
trivial state $\left\vert g\right\rangle =\bigotimes_{j=1}^{2L}\left\vert
+\right\rangle $, which is the ground state of $\mathcal{H}_{\text{X}}$. In
general, the string-order parameter is not quantized.

\subsection{Open chain and edge states}

We next consider an open chain whose edge sites are $j=1$ and $j=2L$. The
stabilizers at the edges are given by%
\begin{align}
K_{1}^{\text{ZXZ}}=& U_{\text{CZ}}\text{X}_{1}U_{\text{CZ}}^{\dagger }=\text{%
X}_{1}\text{Z}_{2},  \label{K1} \\
K_{2L}^{\text{ZXZ}}=& U_{\text{CZ}}\text{X}_{2L}U_{\text{CZ}}^{\dagger }=%
\text{Z}_{2L-1}\text{X}_{2L},  \label{K2L}
\end{align}%
while those at the bulk are given by\cite{Got}%
\begin{equation}
K_{j}^{\text{ZXZ}}=\text{Z}_{j-1}\text{X}_{j}\text{Z}_{j+1}  \label{Kj}
\end{equation}%
for $2\leq j\leq 2L-1$. The energy spectrum is identical to that of the
Hamiltonian (\ref{HX}), because the Hamiltonian is unitary equivalent to the
trivial Hamiltonian (\ref{HX}).

In order to see that the ZXZ model has a symmetry protected topological
phase, we project the Hamiltonian (\ref{HZXZOpen}) onto the $\mathbb{Z}_{2}^{%
\text{even}}\times \mathbb{Z}_{2}^{\text{odd}}$ symmetric space,%
\begin{equation}
\widetilde{\mathcal{H}}_{\text{ZXZ}}=P_{\text{ZXZ}}\mathcal{H}_{\text{ZXZ}%
}P_{\text{ZXZ}}=-\sum_{j=2}^{2L-1}K_{j}^{\text{ZXZ}},
\end{equation}%
where we have used the relation%
\begin{equation}
P_{\text{ZXZ}}K_{1}^{\text{ZXZ}}P_{\text{ZXZ}}=P_{\text{ZXZ}}K_{2L}^{\text{%
ZXZ}}P_{\text{ZXZ}}=0.
\end{equation}%
The stabilizers at the edges $K_{1}^{\text{ZXZ}}$ and $K_{2L}^{\text{ZXZ}}$
are absent in $\widetilde{\mathcal{H}}_{\text{ZXZ}}$. The eigenspectrum for
an open chain is shown in Fig.\ref{FigZXZEne}(b), where there are four-fold
degenerate gapped ground state as indicated by a closed blue oval.

It is understood as follows. We make a unitary transformation by the CZ gate,%
\begin{equation}
\text{CZ}\widetilde{\mathcal{H}}_{\text{ZXZ}}\text{CZ}=-\sum_{j=2}^{2L-1}%
\text{X}_{j}.
\end{equation}%
There is no operator for the qubits with $j=1$ and $2L$ in the Hamiltonian.
Then, the ground state is%
\begin{equation}
\left\vert \psi \right\rangle =\left\vert s_{1}\right\rangle \otimes \left(
\bigotimes_{j=2}^{2L-1}\left\vert +\right\rangle \right) \otimes \left\vert
s_{2L}\right\rangle ,
\end{equation}%
where we have used%
\begin{equation}
\text{X}_{j}\left\vert +\right\rangle =\left\vert +\right\rangle
\end{equation}%
for $2\leq j\leq 2L-1$.

Logical qubits are determined as follows. i) The operators commute with all
the stabilizers $K_{j}^{\text{ZXZ}}$, which means that they are centralizers
of the group spanned by the stabilizers. ii) They are linearly independent
of the stabilizers. iii) They satisfy the Pauli group.

They are explicitly given by 
\begin{align}
\text{X}_{\text{left}}\equiv & \text{CZ}_{1,2}\text{X}_{1}\text{CZ}_{1,2}=%
\text{X}_{1}\text{Z}_{2},\quad \\
\text{Y}_{\text{left}}\equiv & \text{CZ}_{1,2}\text{Y}_{1}\text{CZ}_{1,2}=%
\text{Y}_{1}\text{Z}_{2},\quad \\
\text{Z}_{\text{left}}\equiv & \text{CZ}_{1,2}\text{Z}_{1}\text{CZ}_{1,2}=%
\text{Z}_{1}
\end{align}%
for the left edge, and 
\begin{align}
\text{X}_{\text{right}}\equiv & \text{CZ}_{2L-1,2L}\text{X}_{2L}\text{CZ}%
_{2L-1,2L}=\text{X}_{2L}\text{Z}_{2L-1},\quad \\
\text{Y}_{\text{right}}\equiv & \text{CZ}_{2L-1,2L}\text{Y}_{2L}\text{CZ}%
_{2L-1,2L}=\text{Y}_{2L}\text{Z}_{2L-1},\quad \\
\text{Z}_{\text{right}}\equiv & \text{CZ}_{2L-1,2L}\text{Z}_{2L}\text{CZ}%
_{2L-1,2L}=\text{Z}_{2L}
\end{align}%
for the right edge, where we have used the relations%
\begin{align}
\text{CZ}_{j,j+1}\text{X}_{j}\text{CZ}_{j,j+1}=& \text{X}_{j}\text{Z}_{j+1},
\\
\text{CZ}_{j,j+1}\text{X}_{j+1}\text{CZ}_{j,j+1}=& \text{Z}_{j}\text{X}%
_{j+1}.
\end{align}%
They satisfy the SU(2) commutation relation of the spin.

The $\mathbb{Z}_{2}^{\text{even}}\times \mathbb{Z}_{2}^{\text{odd}}$
symmetry is broken at the edge $j=1$ and $2L$, 
\begin{equation}
\left[ \text{X}_{\text{left}},\eta \right] \neq 0,\qquad \left[ \text{X}_{%
\text{right}},\eta \right] \neq 0.
\end{equation}%
It is known as the t'Hooft anomaly\cite{tH}.

\subsection{Second cohomology group}

\label{ApZXZco}

The generators of the $\mathbb{Z}_{2}^{\text{even}}\times \mathbb{Z}_{2}^{%
\text{odd}}$ symmetry (\ref{etaZXZ}) form a group $G$, which have four
components,%
\begin{equation}
G=\left\{ e,\eta _{\text{odd}},\eta _{\text{even}},\eta _{\text{odd}}\eta _{%
\text{even}}\right\} .
\end{equation}%
The one-dimensional SPT phase is classified by the second cohomology group%
\cite{Chen,Else},%
\begin{equation}
H^{2}\left( G,U\left( 1\right) \right) =\mathbb{Z}_{2}
\end{equation}%
for $G=\mathbb{Z}_{2}\times \mathbb{Z}_{2}$. The 2-cochain is defined by%
\begin{equation}
\omega :G\times G\rightarrow U\left( 1\right) .
\end{equation}%
It is a 2-cocycle if the condition%
\begin{equation}
\omega \left( g,h\right) \omega \left( gh,k\right) =\omega \left( h,k\right)
\omega \left( g,hk\right)
\end{equation}%
is satisfied. The projective representation is characterized by $\omega
\left( g,h\right) $ defined by%
\begin{equation}
U\left( g\right) U\left( h\right) =\omega \left( g,h\right) U\left(
gh\right) .
\end{equation}%
In usual representation of the group, the representation is linear
representation,%
\begin{equation}
U\left( g\right) U\left( h\right) =U\left( gh\right) ,
\end{equation}%
where $\omega \left( h,k\right) =1$. It corresponds to the trivial
representation. On the other hand, $\omega \left( h,k\right) =-1$\ at the
edge, which corresponds to the topological representation. The cocycle
condition is automatically satisfied if the associated law is satisfied
because 
\begin{align}
U\left( g\right) U\left( h\right) U\left( k\right) =& \omega \left(
g,h\right) U\left( gh\right) U\left( k\right)  \notag \\
=& \omega \left( g,h\right) \omega \left( gh,k\right) U\left( ghk\right)
\end{align}%
and%
\begin{align}
U\left( g\right) U\left( h\right) U\left( k\right) =& U\left( g\right)
\omega \left( h,k\right) U\left( hk\right)  \notag \\
=& \omega \left( g,hk\right) \omega \left( h,k\right) U\left( ghk\right)
\end{align}%
lead to the cocycle condition%
\begin{equation}
\omega \left( g,h\right) \omega \left( gh,k\right) =\omega \left(
g,hk\right) \omega \left( h,k\right) .
\end{equation}%
The explicit representation of $\omega $ is determined as follows. The
projective representations of the edge states are given by

\begin{align}
U_{\text{left}}\left( \eta _{\text{ZXZ}}^{\text{odd}}\right) =& \text{X}_{%
\text{left}},\qquad U_{\text{left}}\left( \eta _{\text{ZXZ}}^{\text{even}%
}\right) =\text{Z}_{\text{left}}, \\
U_{\text{right}}\left( \eta _{\text{ZXZ}}^{\text{odd}}\right) =& \text{Z}_{%
\text{right}},\qquad U_{\text{right}}\left( \eta _{\text{ZXZ}}^{\text{even}%
}\right) =\text{X}_{\text{right}},
\end{align}%
as shown in Appendix \ref{ApZXZpro}.

The edge states have a projective representation,%
\begin{align}
U_{\text{left}}\left( \eta _{\text{ZXZ}}^{\text{odd}}\right) U_{\text{left}%
}\left( \eta _{\text{ZXZ}}^{\text{even}}\right) =& -U_{\text{left}}\left(
\eta _{\text{ZXZ}}^{\text{even}}\right) U_{\text{left}}\left( \eta _{\text{%
ZXZ}}^{\text{odd}}\right) , \\
U_{\text{right}}\left( \eta _{\text{ZXZ}}^{\text{odd}}\right) U_{\text{right}%
}\left( \eta _{\text{ZXZ}}^{\text{even}}\right) =& -U_{\text{right}}\left(
\eta _{\text{ZXZ}}^{\text{even}}\right) U_{\text{right}}\left( \eta _{\text{%
ZXZ}}^{\text{odd}}\right) ,
\end{align}%
with%
\begin{equation}
\omega \left( \eta _{\text{ZXZ}}^{\text{odd}},\eta _{\text{ZXZ}}^{\text{even}%
}\right) =-1.
\end{equation}%
On the other hand, all other pair give a trivial result%
\begin{equation}
\omega =1.
\end{equation}

\subsection{Projective representation of the ZXZ model}

\label{ApZXZpro}

We summarize explicit projective representations of the edge states. The
action of the $\mathbb{Z}_{\text{ZXZ}}^{\text{even}}$ symmetry to the left
logical qubits reads 
\begin{align}
\eta _{\text{ZXZ}}^{\text{even}}\text{X}_{\text{left}}\eta _{\text{ZXZ}}^{%
\text{even}}=& \left( \prod\limits_{j\in \text{even}}\text{X}_{j}\right) 
\text{X}_{1}\text{Z}_{2}\left( \prod\limits_{j\in \text{even}}\text{X}%
_{j}\right)  \notag \\
=& \text{X}_{1}\text{X}_{2}\text{Z}_{2}\text{X}_{2}=-\text{X}_{1}\text{Z}%
_{2}=-\text{X}_{\text{left}}
\end{align}%
and%
\begin{align}
\eta _{\text{ZXZ}}^{\text{even}}\text{Z}_{\text{left}}\eta _{\text{ZXZ}}^{%
\text{even}}=& \left( \prod\limits_{j\in \text{even}}\text{X}_{j}\right) 
\text{Z}_{1}\left( \prod\limits_{j\in \text{even}}\text{X}_{j}\right) 
\notag \\
=& \text{Z}_{1}=\text{Z}_{\text{left}},
\end{align}%
Then, the representation of the $\mathbb{Z}_{\text{ZXZ}}^{\text{even}}$
symmetry at the left edge is given by 
\begin{equation}
U_{\text{left}}\left( \eta _{\text{ZXZ}}^{\text{even}}\right) =\text{Z}_{%
\text{left}}.
\end{equation}%
Similarly, the action of the $\mathbb{Z}_{\text{ZXZ}}^{\text{odd}}$ symmetry
to the left logical qubits reads 
\begin{align}
\eta _{\text{ZXZ}}^{\text{odd}}\text{X}_{\text{left}}\eta _{\text{ZXZ}}^{%
\text{odd}}& =\left( \prod\limits_{j\in \text{odd}}\text{X}_{j}\right) 
\text{X}_{1}\text{Z}_{2}\left( \prod\limits_{j\in \text{odd}}\text{X}%
_{j}\right)  \notag \\
=& \text{X}_{1}\text{Z}_{2}=\text{X}_{\text{left}}
\end{align}%
and%
\begin{align}
\eta _{\text{ZXZ}}^{\text{odd}}\text{Z}_{\text{left}}\eta _{\text{ZXZ}}^{%
\text{odd}}=& \left( \prod\limits_{j\in \text{odd}}\text{X}_{j}\right) 
\text{Z}_{1}\text{Z}_{2}\left( \prod\limits_{j\in \text{odd}}\text{X}%
_{j}\right)  \notag \\
=& \text{X}_{1}\text{Z}_{1}\text{X}_{1}\text{Z}_{2}=-\text{Z}_{1}\text{Z}%
_{2}=-\text{Z}_{\text{left}}.
\end{align}%
Then, the representation of the $\mathbb{Z}_{\text{ZXZ}}^{\text{odd}}$
symmetry at the left edge is given by%
\begin{equation}
U_{\text{left}}\left( \eta _{\text{ZXZ}}^{\text{odd}}\right) =\text{X}_{%
\text{left}}.
\end{equation}%
The action of the $\mathbb{Z}_{\text{ZXZ}}^{\text{even}}$ symmetry to the
right logical qubits reads%
\begin{align}
\eta _{\text{ZXZ}}^{\text{even}}\text{X}_{\text{right}}^{1}\eta _{\text{ZXZ}%
}^{\text{even}}=& \left( \prod\limits_{j\in \text{even}}\text{X}_{j}\right) 
\text{X}_{2L}\text{Z}_{2L-1}\left( \prod\limits_{j\in \text{even}}\text{X}%
_{j}\right)  \notag \\
=& \text{X}_{2L}\text{X}_{2L}\text{Z}_{2L-1}\text{X}_{2L}=\text{X}_{2L}\text{%
Z}_{2L-1}=\text{X}_{\text{left}}^{1}
\end{align}%
and%
\begin{align}
\eta _{\text{ZXZ}}^{\text{even}}\text{Z}_{\text{right}}^{1}\eta _{\text{ZXZ}%
}^{\text{even}}=& \left( \prod\limits_{j\in \text{even}}\text{X}_{j}\right) 
\text{Z}_{2L}\left( \prod\limits_{j\in \text{even}}\text{X}_{j}\right) 
\notag \\
=& \text{X}_{2L}\text{Z}_{2L}\text{X}_{2L}=-\text{X}_{2L}\text{Z}_{2L}\text{X%
}_{2L}=-\text{Z}_{\text{right}}^{1}.
\end{align}%
Then, the representation of the $\mathbb{Z}_{\text{ZXZ}}^{\text{even}}$
symmetry at the right edge is given by%
\begin{equation}
U_{\text{right}}\left( \eta _{\text{ZXZ}}^{\text{even}}\right) =\text{X}_{%
\text{right}}.
\end{equation}%
The action of the $\mathbb{Z}_{\text{ZXZ}}^{\text{odd}}$ symmetry to the
right logical qubits reads%
\begin{align}
\eta _{\text{ZXZ}}^{\text{odd}}\text{X}_{\text{right}}^{1}\eta _{\text{ZXZ}%
}^{\text{odd}}=& \left( \prod\limits_{j\in \text{odd}}\text{X}_{j}\right) 
\text{X}_{2L}\text{Z}_{2L-1}\left( \prod\limits_{j\in \text{odd}}\text{X}%
_{j}\right)  \notag \\
=& \text{X}_{2L-1}\text{X}_{2L}\text{Z}_{2L-1}\text{X}_{2L-1}=\text{X}_{2L}%
\text{Z}_{2L-1}  \notag \\
=& -\text{X}_{\text{right}}^{1}
\end{align}%
and%
\begin{align}
\eta _{\text{ZXZ}}^{\text{odd}}\text{Z}_{\text{right}}^{1}\eta _{\text{ZXZ}%
}^{\text{odd}}=& \left( \prod\limits_{j\in \text{odd}}\text{X}_{j}\right) 
\text{Z}_{2L}\left( \prod\limits_{j\in \text{odd}}\text{X}_{j}\right) 
\notag \\
=& \text{X}_{2L-1}\text{Z}_{2L}\text{X}_{2L-1}=\text{Z}_{2L}=\text{Z}_{\text{%
right}}^{1}.
\end{align}%
Then, the representation of the $\mathbb{Z}_{\text{ZXZ}}^{\text{odd}}$
symmetry at the right edge is given by%
\begin{equation}
U_{\text{right}}\left( \eta _{\text{ZXZ}}^{\text{odd}}\right) =\text{Z}_{%
\text{right}}.
\end{equation}

\begin{figure}[t]
\centerline{\includegraphics[width=0.48\textwidth]{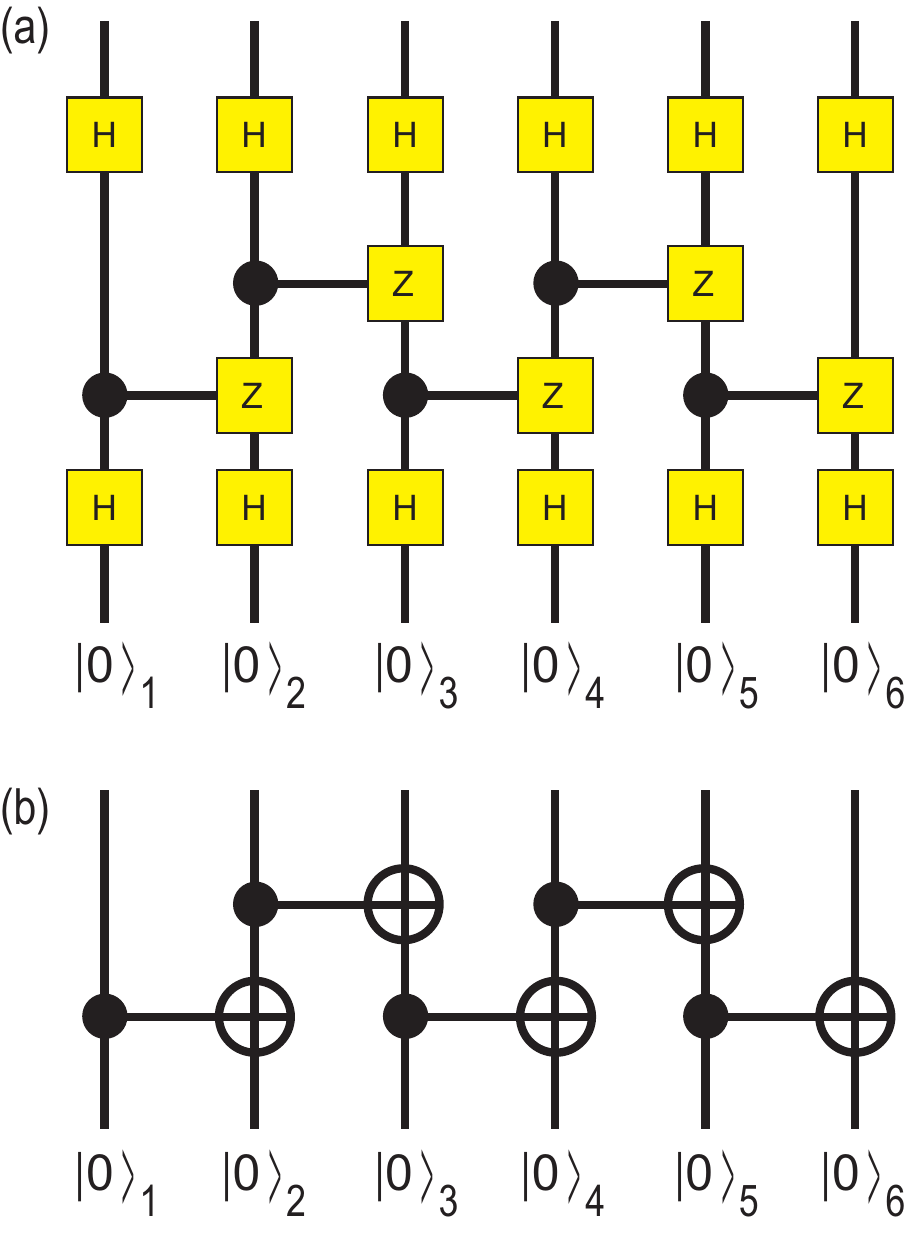}}
\caption{XZX model. (a) Quantum circuit representation. (b) Equivalent
quantum circuit using the CX gates.}
\label{FigXZX}
\end{figure}

\section{XZX model}

\label{ApXZX}

We next study the XZX model given by\cite{Verr,Sadaf}%
\begin{equation}
\mathcal{H}_{\text{XZX}}=-\sum_{j=1}^{2L-1}K_{j}^{\text{XZX}}
\end{equation}%
with the stabilizer%
\begin{equation}
K_{j}^{\text{XZX}}\equiv \text{X}_{j-1}\text{Z}_{j}\text{X}_{j+1}.
\end{equation}%
The Hamiltonian is constructed by applying the Hadamard gates H$_{j}$ to the
ZXZ model (\ref{HZXZK}),%
\begin{equation}
\mathcal{H}_{\text{XZX}}=\left( \prod_{j=1}^{L}\text{H}_{j}\right) \mathcal{%
H}_{\text{ZXZ}}\left( \prod_{j=1}^{L}\text{H}_{j}\right) .
\end{equation}%
The associated cluster state is given by%
\begin{equation}
\left\vert \psi ^{\text{XZX}}\right\rangle =\prod_{j=1}^{L}\text{H}%
_{j}\prod_{j=1}^{2L-1}\text{CZ}_{j,j+1}\bigotimes_{j=1}^{L}\left\vert
+\right\rangle ,
\end{equation}%
which corresponds to Eq.(\ref{UjN}) with 
\begin{equation}
\prod_{j=1}^{2L-N}U_{\left\{ j;N\right\} }=\prod_{j=1}^{L}\text{H}%
_{j}\prod_{j=1}^{2L-1}\text{CZ}_{j,j+1}.
\end{equation}%
The quantum circuit representation is shown in Fig.\ref{FigXZX}(a). It is
also constructed by simply applying the CNOT gate CX$_{j,j+1}$, whose
control qubit is $j$ and target qubit is $j+1$, 
\begin{equation}
\left\vert \psi ^{\text{XZX}}\right\rangle =\prod_{j=1}^{2L-1}\text{CX}%
_{j,j+1}\bigotimes_{j=1}^{L}\left\vert 0\right\rangle .
\end{equation}%
The quantum circuit representation is shown in Fig.\ref{FigXZX}(b).

The generators of the $\mathbb{Z}_{2}^{\text{even}}\times \mathbb{Z}_{2}^{%
\text{odd}}$ symmetry are given by%
\begin{equation}
\eta _{\text{XZX}}^{\text{even}}\equiv \prod\limits_{j\in \text{even}}\text{%
Z}_{j},\qquad \eta _{\text{XZX}}^{\text{odd}}\equiv \prod\limits_{j\in 
\text{odd}}\text{Z}_{j}.
\end{equation}%
The generators of the non-invertible symmetry is%
\begin{equation}
\mathsf{D}_{\text{XZX}}=\mathsf{TD}_{\text{XZX}}^{\text{even}}\mathsf{D}_{%
\text{XZX}}^{\text{odd}},
\end{equation}%
where%
\begin{align}
\mathsf{D}_{\text{XZX}}^{\text{even}}=& e^{\frac{\pi iN}{8}}\frac{1+\eta _{%
\text{XZX}}}{2}e^{-\frac{i\pi }{4}\text{Z}_{2L}}\mathsf{D}_{2L-2}^{\text{XZX}%
}\cdots \mathsf{D}_{4}^{\text{XZX}}\mathsf{D}_{2}^{\text{XZX}}, \\
\mathsf{D}_{\text{XZX}}^{\text{odd}}=& e^{\frac{\pi iN}{8}}\frac{1+\eta _{%
\text{XZX}}}{2}e^{-\frac{i\pi }{4}\text{Z}_{2L-1}}\mathsf{D}_{2L-3}^{\text{%
XZX}}\cdots \mathsf{D}_{3}^{\text{XZX}}\mathsf{D}_{1}^{\text{XZX}}
\end{align}%
with%
\begin{equation}
\mathsf{D}_{j}^{\text{XZX}}\equiv e^{-\frac{i\pi }{4}\text{X}_{j}\text{X}%
_{j+1}}e^{-\frac{i\pi }{4}\text{Z}_{j}}
\end{equation}%
and%
\begin{equation}
\eta _{\text{XZX}}\equiv \eta _{\text{XZX}}^{\text{even}}\eta _{\text{XZX}}^{%
\text{odd}}.
\end{equation}%
The action of the non-invertible symmetry $\mathsf{D}_{\text{XZX}}$ is%
\begin{equation}
\text{Z}_{j}\rightsquigarrow \text{X}_{j-1}\text{X}_{j+1},\qquad \text{X}%
_{j-1}\text{X}_{j+1}\rightsquigarrow \text{Z}_{j}.
\end{equation}%
The action of the KT transformation is%
\begin{align}
\text{Z}_{j}\leadsto & \text{Z}_{j}, \\
\text{X}_{j-1}\text{Z}_{j}\text{X}_{j+1}\leadsto & \text{X}_{j-1}\text{X}%
_{j+1}.
\end{align}%
The string-order parameter is given by%
\begin{align}
\mathcal{O}\left( i,j\right) \equiv & \left\langle g\right\vert
\prod\limits_{k=i}^{j}K_{k}^{\text{XZX}}\left\vert g\right\rangle  \notag \\
=& \left\langle g\right\vert \text{X}_{i}\left( \prod\limits_{k=i+1}^{j-1}%
\text{Z}_{k}\right) \text{X}_{j}\left\vert g\right\rangle .
\end{align}

\begin{figure}[t]
\centerline{\includegraphics[width=0.48\textwidth]{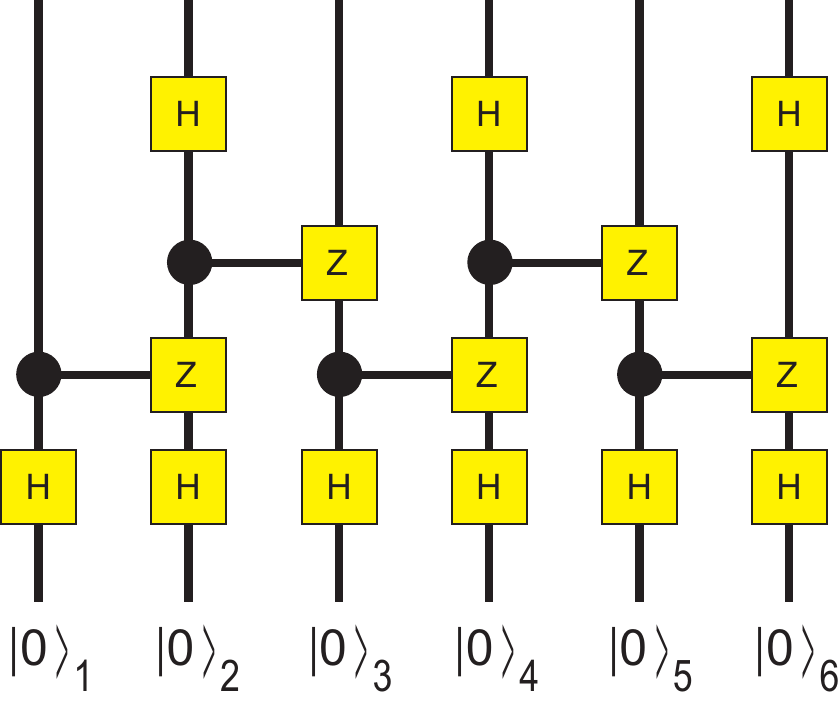}}
\caption{ZZZ-XXX model. Quantum circuit representation.}
\label{FigZZZ}
\end{figure}

\section{ZZZ-XXX model}

\label{ApZZZ}

The ZZZ-XXX model is given by\cite{Jia,Jia2}%
\begin{equation}
\mathcal{H}_{\text{ZZZ}}=-\sum_{j=1}^{L-1}\left( K_{2j}^{\text{ZZZ}%
}+K_{2j+1}^{\text{XXX}}\right)
\end{equation}%
with the stabilizers 
\begin{align}
K_{2j}^{\text{ZZZ}}\equiv & \text{Z}_{2j-1}\text{Z}_{2j}\text{Z}_{2j+1}, \\
K_{2j+1}^{\text{XXX}}\equiv & \text{X}_{2j}\text{X}_{2j+1}\text{X}_{2j+2}.
\end{align}%
It is constructed by applying the Hadamard gates only to the even qubits in
the ZXZ model,%
\begin{equation}
\mathcal{H}_{\text{ZZZ}}=\left( \prod_{j=1}^{L}\text{H}_{2j}\right) 
\mathcal{H}_{\text{ZXZ}}\prod_{j=1}^{L}\text{H}_{2j}.
\end{equation}%
The associated cluster state is given by%
\begin{equation}
\left\vert \psi ^{\text{ZZZ}}\right\rangle =\prod_{j=1}^{L}\text{H}%
_{2j}\prod_{j=1}^{2L-1}\text{CZ}_{j,j+1}\bigotimes_{j=1}^{L}\left\vert
+\right\rangle ,
\end{equation}%
which corresponds to Eq.(\ref{UjN}) with%
\begin{equation}
\prod_{j=1}^{2L-N}U_{\left\{ j;N\right\} }=\prod_{j=1}^{L}\text{H}%
_{2j}\prod_{j=1}^{2L-1}\text{CZ}_{j,j+1},
\end{equation}%
where%
\begin{align}
U_{\left\{ 2j;0\right\} }=& \text{H}_{2j}, \\
U_{\left\{ 2j-1;0\right\} }=& I_{2}, \\
U_{\left\{ j;1\right\} }=& \text{CZ}_{j,j+1}.
\end{align}%
The quantum circuit representation is shown in Fig.\ref{FigZZZ}.

The generators of the $\mathbb{Z}_{2}^{\text{even}}\times \mathbb{Z}_{2}^{%
\text{odd}}$ symmetry are given by%
\begin{equation}
\eta _{\text{ZZZ}}^{\text{even}}\equiv \prod\limits_{j\in \text{even}}\text{%
X}_{j},\qquad \eta _{\text{ZZZ}}^{\text{odd}}\equiv \prod\limits_{j\in 
\text{odd}}\text{Z}_{j}.
\end{equation}%
The generator of the non-invertible symmetry is%
\begin{equation}
\mathsf{D}_{\text{ZZZ}}=\mathsf{TD}_{\text{ZZZ}}^{\text{even}}\mathsf{D}_{%
\text{ZZZ}}^{\text{odd}},
\end{equation}%
where%
\begin{align}
\mathsf{D}_{\text{ZZZ}}^{\text{even}}=& e^{\frac{\pi iN}{8}}\frac{1+\eta _{%
\text{ZZZ}}}{2}e^{-\frac{i\pi }{4}\text{Z}_{2L}}\mathsf{D}_{2L-2}^{\text{ZZZ}%
}\cdots \mathsf{D}_{4}^{\text{ZZZ}}\mathsf{D}_{2}^{\text{ZZZ}}, \\
\mathsf{D}_{\text{ZZZ}}^{\text{odd}}=& e^{\frac{\pi iN}{8}}\frac{1+\eta _{%
\text{ZZZ}}}{2}e^{-\frac{i\pi }{4}\text{X}_{2L-1}}\mathsf{D}_{2L-2}^{\text{%
ZZZ}}\cdots \mathsf{D}_{3}^{\text{ZZZ}}\mathsf{D}_{1}^{\text{ZZZ}}
\end{align}%
with%
\begin{align}
\mathsf{D}_{2j}^{\text{ZZZ}}\equiv & e^{-\frac{i\pi }{4}\text{X}_{2j}\text{Z}%
_{2j+1}}e^{-\frac{i\pi }{4}\text{Z}_{2j}}, \\
\mathsf{D}_{2j-1}^{\text{ZZZ}}\equiv & e^{-\frac{i\pi }{4}\text{Z}_{2j-1}%
\text{X}_{j}}e^{-\frac{i\pi }{4}\text{X}_{2j-1}}
\end{align}%
and%
\begin{equation}
\eta _{\text{ZZZ}}\equiv \eta _{\text{ZZZ}}^{\text{even}}\eta _{\text{ZZZ}}^{%
\text{odd}}.
\end{equation}%
The action of the non-invertible symmetry $\mathsf{D}_{\text{ZZZ}}$ is%
\begin{align}
\text{X}_{2j-1}& \rightsquigarrow \text{Z}_{2j-2}\text{Z}_{2j},\qquad \text{Z%
}_{2j-2}\text{Z}_{2j}\rightsquigarrow \text{X}_{2j-1},  \notag \\
\text{Z}_{2j}\rightsquigarrow & \text{X}_{2j-1}\text{X}_{2j+1},\qquad \text{X%
}_{2j-1}\text{X}_{2j+1}\rightsquigarrow \text{Z}_{2j}.
\end{align}%
The action of the KT transformation is%
\begin{align}
\text{X}_{2j-1}\leadsto & \text{X}_{2j-1},\qquad \text{Z}_{2j}\leadsto \text{%
Z}_{2j}, \\
\text{Z}_{2j-1}\text{Z}_{2j}\text{Z}_{2j+1}\leadsto & \text{Z}_{2j-1}\text{Z}%
_{2j+1}, \\
\text{X}_{2j}\text{X}_{2j+1}\text{X}_{2j+2}\leadsto & \text{X}_{2j}\text{X}%
_{2j+2}.
\end{align}%
The string-order parameter is given by%
\begin{equation}
\mathcal{O}\left( i,j\right) =\left\langle g\right\vert
\prod\limits_{k=i}^{j}K_{k}\left\vert g\right\rangle .
\end{equation}%
It is explicitly given by%
\begin{equation}
\mathcal{O}\left( i,j\right) =\left\langle g\right\vert \text{Z}_{i}\left(
\prod\limits_{k=i+1}^{j-1}i\text{Y}_{k}\right) \text{Z}_{j}\left\vert
g\right\rangle
\end{equation}%
for odd $i$ and $j$,%
\begin{equation}
\mathcal{O}\left( i,j\right) =\left\langle g\right\vert \text{X}_{i}\left(
\prod\limits_{k=i+1}^{j-1}i\text{Y}_{k}\right) \text{X}_{j}\left\vert
g\right\rangle
\end{equation}%
for even $i$ and $j$,%
\begin{equation}
\mathcal{O}\left( i,j\right) =\left\langle g\right\vert \text{Z}_{i}\left(
\prod\limits_{k=i+1}^{j-1}i\text{Y}_{k}\right) \text{X}_{j}\left\vert
g\right\rangle
\end{equation}%
for odd $i$ and even $j$,%
\begin{equation}
\mathcal{O}\left( i,j\right) =\left\langle C\right\vert \text{X}_{i}\left(
\prod\limits_{k=i+1}^{j-1}i\text{Y}_{k}\right) \text{Z}_{j}\left\vert
C\right\rangle
\end{equation}%
for even $i$ and odd $j$.

\begin{figure}[t]
\centerline{\includegraphics[width=0.48\textwidth]{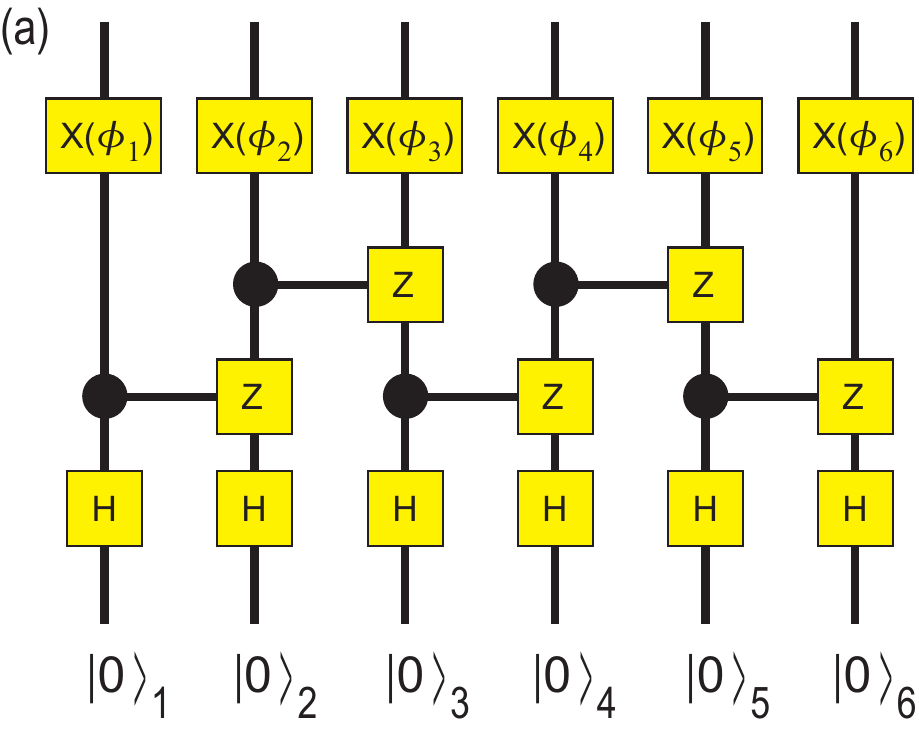}}
\caption{Bit-flip model. (a) Quantum circuit generating the bit-flipped ZXZ
cluster state.}
\label{FigBit}
\end{figure}

\section{Bit flip model}

\label{ApBit}

In actual quantum computer, there is a bit flip error. We model it by the X
rotation with an arbitrary angle $\phi _{j}$ depending on the site $j$,%
\begin{equation}
U_{\left\{ j;N\right\} }=e^{-i\frac{\phi _{j}}{2}\text{X}_{j}}\equiv U_{j}^{%
\text{bit}},
\end{equation}%
as corresponds to Eq.(\ref{UjN}) with $N=0$. It is not a Clifford gate. Its
action on Z$_{j}$ is%
\begin{equation}
U_{j}^{\text{bit}}\text{Z}_{j}\left( U_{j}^{\text{bit}}\right) ^{-1}=\text{Z}%
_{j}\cos \phi _{j}-\text{Y}_{j}\sin \phi _{j}\equiv \text{Z}_{j}\left( \phi
\right) .
\end{equation}%
The generated cluster state is%
\begin{equation}
\left\vert \psi ^{\text{bit}}\right\rangle =\prod_{j=1}^{2L}U_{j}^{\text{bit%
}}\prod_{j=1}^{2L-1}\text{CZ}_{j,j+1}\bigotimes_{j=1}^{L}\left\vert
+\right\rangle .
\end{equation}%
The quantum circuit representation is shown in Fig.\ref{FigBit}. We note
that it is not the IQP because $U_{j}^{\text{bit}}$\ is not a diagonal gate.

The corresponding Hamiltonian is%
\begin{equation}
\mathcal{H}_{\text{bit}}=-\sum_{j}\text{Z}_{j-1}\left( \phi \right) \text{X}%
_{j}\text{Z}_{j+1}\left( \phi \right) .
\end{equation}%
The generators of the $\mathbb{Z}_{2}^{\text{even}}\times \mathbb{Z}_{2}^{%
\text{odd}}$ symmetry are identical to those of the ZXZ model. The generator
of the non-invertible symmetry is%
\begin{equation}
\mathsf{D}_{\text{bit}}=\mathsf{TD}_{\text{bit}}^{\text{even}}\mathsf{D}_{%
\text{bit}}^{\text{odd}},
\end{equation}%
where%
\begin{align}
\mathsf{D}_{\text{bit}}^{\text{even}}=& e^{\frac{\pi iN}{8}}\frac{1+\eta _{%
\text{ZXZ}}}{2}e^{-\frac{i\pi }{4}\text{X}_{2L}}\mathsf{D}_{2L-2}^{\text{bit}%
}\cdots \mathsf{D}_{4}^{\text{bit}}\mathsf{D}_{2}^{\text{bit}}, \\
\mathsf{D}_{\text{bit}}^{\text{odd}}=& e^{\frac{\pi iN}{8}}\frac{1+\eta _{%
\text{ZXZ}}}{2}e^{-\frac{i\pi }{4}\text{X}_{2L-1}}\mathsf{D}_{2L-3}^{\text{%
bit}}\cdots \mathsf{D}_{3}^{\text{bit}}\mathsf{D}_{1}^{\text{bit}}
\end{align}%
with Eq.(\ref{etaZXZ2}) and 
\begin{equation}
\mathsf{D}_{j}^{\text{bit}}\equiv e^{-\frac{i\pi }{4}\text{Z}_{j}\left( \phi
\right) \text{Z}_{j+1}\left( \phi \right) }e^{-\frac{i\pi }{4}\text{X}_{j}}.
\end{equation}%
The action of the non-invertible symmetry $\mathsf{D}_{\text{bit}}$ is%
\begin{align}
\text{X}_{j} \rightsquigarrow &\text{Z}_{j-1}\left( \phi \right) \text{Z}%
_{j+1}\left( \phi \right) ,\qquad \\
\text{Z}_{j-1}\left( \phi \right) \text{Z}_{j+1}\left( \phi \right)
\rightsquigarrow &\text{X}_{j}.
\end{align}%
The four-fold degenerate ground states are robust for an open chain even
though we have introduced the site-dependent bit flip.

The action of the KT transformation is%
\begin{align}
\text{X}_{j}\leadsto & \text{X}_{j}, \\
\text{Z}_{j-1}\left( \phi \right) \text{X}_{j}\text{Z}_{j+1}\left( \phi
\right) \leadsto & \text{Z}_{j-1}\left( \phi \right) \text{Z}_{j+1}\left(
\phi \right) .
\end{align}%
The string-order parameter is given by%
\begin{align}
\mathcal{O}\left( i,j\right) \equiv & \left\langle g\right\vert
\prod\limits_{k=i}^{j}K_{k}^{\text{Bit}}\left\vert g\right\rangle  \notag \\
=& \left\langle g\right\vert \text{Z}_{i}\left( \phi \right) \left(
\prod\limits_{k=i+1}^{j-1}\text{X}_{k}\right) \text{Z}_{j}\left( \phi
\right) \left\vert g\right\rangle .
\end{align}

\begin{figure}[t]
\centerline{\includegraphics[width=0.48\textwidth]{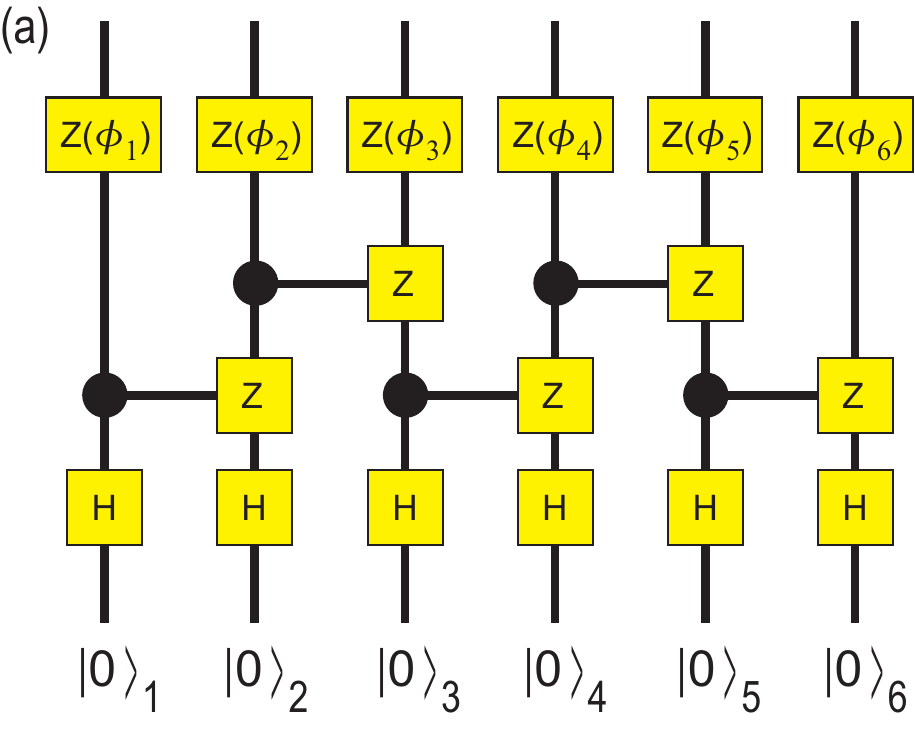}}
\caption{Phase-flip model. (a) Quantum circuit generating the phase-flipped
ZXZ cluster state.}
\label{FigPhase}
\end{figure}

\section{Phase flip model}

\label{ApPhase}

In addition to the bit flip, there is also a phase flip in actual quantum
computer, which we model by the quantum gate%
\begin{equation}
U_{j}^{\text{phase}}\equiv e^{i\phi _{j}\text{Z}_{j}}.
\end{equation}%
It is not a Clifford gate\ except for $\phi _{j}=0$ and $\pi $. Its action
on X$_{j}$ is%
\begin{equation}
U_{j}^{\text{phase}}\text{X}_{j}\left( U_{j}^{\text{phase}}\right) ^{-1}=%
\text{X}_{j}\cos \phi _{j}-\text{Y}_{j}\sin \phi _{j}\equiv \text{X}%
_{j}\left( \phi _{j}\right) .
\end{equation}%
The generated cluster state is%
\begin{equation}
\left\vert \psi ^{\text{phase}}\right\rangle =\prod_{j=1}^{2L}U_{j}^{\text{%
phase}}\prod_{j=1}^{2L-1}\text{CZ}_{j,j+1}\bigotimes_{j=1}^{L}\left\vert
+\right\rangle .
\end{equation}%
The quantum circuit representation is shown in Fig.\ref{FigPhase}. The
corresponding Hamiltonian is%
\begin{equation}
\mathcal{H}_{\text{phase}}=-\sum_{j}\text{Z}_{j-1}\text{X}_{j}\left( \phi
_{j}\right) \text{Z}_{j+1}.
\end{equation}%
The generators of the $\mathbb{Z}_{2}^{\text{even}}\times \mathbb{Z}_{2}^{%
\text{odd}}$ symmetry are given by%
\begin{equation}
\eta _{\text{phase}}^{\text{even}}\equiv \prod\limits_{j\in \text{even}}%
\text{X}_{j}\left( \phi _{j}\right) ,\qquad \eta _{\text{phase}}^{\text{odd}%
}\equiv \prod\limits_{j\in \text{odd}}\text{X}_{j}\left( \phi _{j}\right) .
\end{equation}%
The generator of the non-invertible symmetry is%
\begin{equation}
\mathsf{D}_{\text{phase}}=\mathsf{TD}_{\text{phase}}^{\text{even}}\mathsf{D}%
_{\text{phase}}^{\text{odd}},
\end{equation}%
where%
\begin{align}
\mathsf{D}_{\text{phase}}^{\text{even}}=& e^{\frac{\pi iN}{8}}\frac{1+\eta _{%
\text{phase}}}{2}e^{-\frac{i\pi }{4}\text{X}_{2L}\left( \phi \right) }%
\mathsf{D}_{2L-2}^{\text{phase}}\cdots \mathsf{D}_{4}^{\text{phase}}\mathsf{D%
}_{2}^{\text{phase}}, \\
\mathsf{D}_{\text{phase}}^{\text{odd}}=& e^{\frac{\pi iN}{8}}\frac{1+\eta _{%
\text{bit}}}{2}e^{-\frac{i\pi }{4}\text{X}_{2L-1}\left( \phi \right) }%
\mathsf{D}_{2L-3}^{\text{phase}}\cdots \mathsf{D}_{3}^{\text{phase}}\mathsf{D%
}_{1}^{\text{phase}}
\end{align}%
with%
\begin{equation}
\mathsf{D}_{j}^{\text{phase}}\equiv e^{-\frac{i\pi }{4}\text{Z}_{j}\text{Z}%
_{j+1}}e^{-\frac{i\pi }{4}\text{X}_{j}\left( \phi \right) }
\end{equation}%
and%
\begin{equation}
\eta _{\text{phase}}\equiv \eta _{\text{phase}}^{\text{even}}\eta _{\text{%
phase}}^{\text{odd}}.
\end{equation}%
The action of the non-invertible symmetry $\mathsf{D}_{\text{phase}}$ is%
\begin{equation}
\text{X}_{j}\left( \phi \right) \rightsquigarrow \text{Z}_{j-1}\text{Z}%
_{j+1},\qquad \text{Z}_{j-1}\text{Z}_{j+1}\rightsquigarrow \text{X}%
_{j}\left( \phi \right) .
\end{equation}%
The four-fold degenerate ground states are robust for an open chain even
though we have introduced the site-dependent phase flip.

The action of the KT transformation is%
\begin{align}
\text{X}_{j}\left( \phi \right) \leadsto & \text{X}_{j}\left( \phi \right) ,
\\
\text{Z}_{j-1}\text{X}_{j}\left( \phi \right) \text{Z}_{j+1}\leadsto & \text{%
Z}_{j-1}\text{Z}_{j+1}.
\end{align}%
The string-order parameter is given by%
\begin{align}
\mathcal{O}\left( i,j\right) \equiv & \left\langle g\right\vert
\prod\limits_{k=i}^{j}K_{k}^{\text{phase}}\left\vert g\right\rangle  \notag
\\
=& \left\langle g\right\vert \text{Z}_{i}\left( \prod\limits_{k=i+1}^{j-1}%
\text{X}_{k}\left( \phi \right) \right) \text{Z}_{j}\left\vert
g\right\rangle .
\end{align}

\begin{figure}[t]
\centerline{\includegraphics[width=0.48\textwidth]{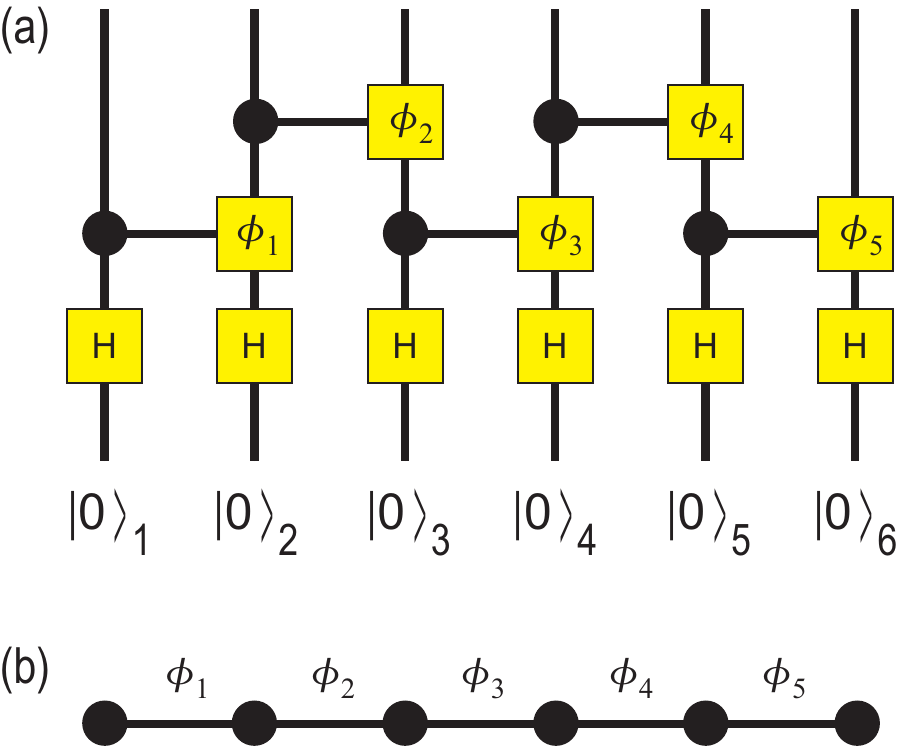}}
\caption{CP gate model. (a) Quantum circuit representation. (b) Graph
representation, where the black disks represent the state $\left\vert
+\right\rangle $ and the segments represent the CP gates.}
\label{FigCR}
\end{figure}

\section{Controlled phase shift gate model}

\label{ApCP}

The CZ gate is actually equipped by time evolution of the CP gate by the
fine tuning of $\phi =\pi $. It is hard to precisely tune the phase shift to
be $\phi =\pi $\ in actual experiments. Hence, it is meaningful to
investigate the cluster state generated by the CP gate.

We generalize the cluster state by using the CP gate instead of the CZ gate,%
\begin{align}
\text{CP}_{j,j+1}\left( \phi _{j}\right) & \equiv e^{-i\frac{\phi _{j}}{4}%
}e^{i\frac{\phi _{j}}{4}\text{Z}_{j}\otimes \text{I}_{j+1}}e^{i\frac{\phi
_{j}}{4}\text{I}_{j}\otimes \text{Z}_{j+1}}U_{\text{ZZ}}\left( \phi
_{j}\right)  \notag \\
& =\text{diag}\left( 1,1,1,e^{i\phi _{j}}\right) _{j,j+1},
\end{align}%
where we have assumed a qubit-dependent angle $\phi _{j}$ and the Ising gate%
\begin{equation}
U_{\text{ZZ}}\left( \phi _{j}\right) \equiv e^{-i\frac{\phi _{j}}{4}\text{Z}%
_{j}\otimes \text{Z}_{j+1}}.
\end{equation}%
See Appendix \ref{ApIsing} for the Ising gate.

It is not a Clifford gate except for $\phi _{j}=0$ and $\pi $. The cluster
state is generated by 
\begin{equation}
\left\vert \text{C}\left( \left\{ \phi _{j}\right\} \right) \right\rangle
\equiv \prod_{j=1}^{2L-1}\text{CP}_{j,j+1}\left( \phi _{j}\right)
\bigotimes_{j=1}^{L}\left\vert +\right\rangle ,
\end{equation}%
which is of the form of Eq.(\ref{UjN}) with $N=1$,%
\begin{equation}
\prod_{j=1}^{2L-1}U_{\left\{ j;1\right\} }=\prod_{j=1}^{2L-1}\text{CP}%
_{j,j+1}\left( \phi _{j}\right) .
\end{equation}%
The quantum circuit representation is shown in Fig.\ref{FigCR}(a). The
generated cluster state is called the weighted graph state\cite{Dur,Hartmann}%
. The weighted graph representation is shown in Fig.\ref{FigCR}(b).

The corresponding Hamiltonian is given by%
\begin{align}
\mathcal{H}_{\text{CP}}=& \sum_{j,\left\{ \zeta _{j}\right\} }\alpha _{\zeta
_{j-1}\zeta _{j+1}}\text{Z}_{j-1}^{\zeta _{j-1}}\text{X}_{j}\text{Z}%
_{j+1}^{\zeta _{j+1}}  \notag \\
& +\sum_{j,\left\{ \upsilon _{j}\right\} }\beta _{\upsilon _{j-1}\upsilon
_{j+1}}\text{Y}_{j-1}^{\upsilon _{j-1}}\text{X}_{j}\text{Y}_{j+1}^{\upsilon
_{j+1}}
\end{align}%
with%
\begin{align}
\alpha _{00}=& \frac{1}{4}\left( 1+\cos \phi _{j}+\cos \phi _{j+1}+\cos
\left( \phi _{j}+\phi _{j+1}\right) \right) , \\
\alpha _{10}=& \frac{1}{4}\left( 1-\cos \phi _{j}+\cos \phi _{j+1}-\cos
\left( \phi _{j}+\phi _{j+1}\right) \right) , \\
\alpha _{01}=& \frac{1}{4}\left( 1+\cos \phi _{j}-\cos \phi _{j+1}-\cos
\left( \phi _{j}+\phi _{j+1}\right) \right) , \\
\alpha _{11}=& \frac{1}{4}\left( 1-\cos \phi _{j}-\cos \phi _{j+1}+\cos
\left( \phi _{j}+\phi _{j+1}\right) \right) , \\
\beta _{00}=& \frac{1}{4}\left( \sin \phi _{j}+\sin \phi _{j+1}+\sin \left(
\phi _{j}+\phi _{j+1}\right) \right) , \\
\beta _{10}=& \frac{1}{4}\left( -\sin \phi _{j}+\sin \phi _{j+1}-\sin \left(
\phi _{j}+\phi _{j+1}\right) \right) , \\
\beta _{01}=& \frac{1}{4}\left( \sin \phi _{j}-\sin \phi _{j+1}-\sin \left(
\phi _{j}+\phi _{j+1}\right) \right) , \\
\beta _{11}=& \frac{1}{4}\left( -\sin \phi _{j}-\sin \phi _{j+1}+\sin \left(
\phi _{j}+\phi _{j+1}\right) \right) ,
\end{align}%
where we have used the relation%
\begin{align}
& \text{CP}_{j,j+1}\left( \phi _{j}\right) \text{X}_{j+1}\text{CP}%
_{j,j+1}\left( \phi _{j}\right)   \notag \\
=& \text{X}_{j+1}\cos \left( \phi _{j}\right) +\text{Y}_{j+1}\sin \left(
\phi _{j}\right) .
\end{align}%
They are simplified for a constant angle $\phi _{j}=\phi $ as%
\begin{align}
\alpha _{00}=& \cos ^{2}\frac{\phi }{2}\cos \phi , \\
\alpha _{10}=& \alpha _{01}=\frac{\sin ^{2}\phi }{2}, \\
\alpha _{11}=& -\sin ^{2}\frac{\phi }{2}\cos \phi , \\
\beta _{00}=& \cos ^{2}\frac{\phi }{2}\sin \phi , \\
\beta _{10}=& \beta _{01}=-\frac{\sin 2\phi }{4}, \\
\beta _{11}=& -\sin ^{2}\frac{\phi }{2}\sin \phi .
\end{align}%
It is interesting that the four-fold degenerate edge states are robust even
for small angle rotation $\phi $. It is a merit for experimental realization
in photonic systems\cite{Yamazaki}, because it is hard to make a large angle
rotation $\phi =\pi $ due to weak interactions between photons.

The action of the non-invertible symmetry is%
\begin{equation}
\widetilde{\text{X}}_{j}\rightsquigarrow \text{Z}_{j-1}\text{Z}_{j+1},\qquad 
\text{Z}_{j-1}\text{Z}_{j+1}\rightsquigarrow \widetilde{\text{X}}_{j}
\end{equation}%
with%
\begin{align}
\widetilde{\text{X}}_{j}\left( \phi \right) \equiv & \text{X}_{j}\cos \phi
\cos ^{2}\frac{\phi }{2}+\text{Y}_{j}\sin \phi \cos ^{2}\frac{\phi }{2} 
\notag \\
& +\left( \text{X}_{j}\text{Z}_{j+1}+\text{Z}_{j-1}\text{X}_{j}\right) \frac{%
\sin ^{2}\phi }{2}  \notag \\
& -\left( \text{Y}_{j}\text{Z}_{j+1}+\text{Z}_{j-1}\text{Y}_{j}\right) \frac{%
\sin 2\phi }{4}  \notag \\
& -\text{Z}_{j-1}\text{X}_{j}\text{Z}_{j+1}\cos \phi \sin ^{2}\frac{\phi }{2}
\notag \\
& -\text{Z}_{j-1}\text{Y}_{j}\text{Z}_{j+1}\sin \phi \sin ^{2}\frac{\phi }{2}%
.
\end{align}

The action of the KT transformation is%
\begin{align}
\widetilde{\text{X}}_{j}\left( \phi \right) \leadsto & \widetilde{\text{X}}%
_{j}\left( \phi \right) , \\
\text{Z}_{j-1}\widetilde{\text{X}}_{j}\left( \phi \right) \text{Z}%
_{j+1}\leadsto & \text{Z}_{j-1}\text{Z}_{j+1}.
\end{align}

\section{C$^{N}$P gate model}

\label{ApCNP}

In the similar way, the Hamiltonian corresponding to the C$^{N}$P gate is
given by%
\begin{equation}
\mathcal{H}_{\text{C}^{N}\text{P}}=-\sum_{j}K_{j}^{\text{C}^{N}\text{P}}
\end{equation}%
with the stabilizer%
\begin{equation}
K_{j}^{\text{C}^{N}\text{P}}=\text{C}^{N}\text{P}_{\left\{ j\right\} }\text{X%
}_{j}\text{C}^{N}\text{P}_{\left\{ j\right\} }
\end{equation}%
where%
\begin{equation}
\text{C}^{N}\text{P}_{\left\{ j\right\} }\equiv \prod\limits_{k=0}^{N}\text{%
C}^{N-1}\text{P}_{\left\{ j;N\right\} },
\end{equation}%
with%
\begin{align}
& \text{C}^{N-1}\text{P}_{j-N+k,\cdots ,j+k}  \notag \\
& =\text{diag}\left( 1,1,\cdots ,1,e^{i\phi _{j}}\right) _{j-N+k,\cdots
,j+k}.
\end{align}%
The cluster state is given by%
\begin{equation}
\left\vert \psi \right\rangle =\prod_{j=1}^{2L-N}\text{C}^{N}\text{P}%
_{\left\{ j;N\right\} }\bigotimes_{j=1}^{2L}\left\vert +\right\rangle _{j},
\end{equation}%
which is of the form of Eq.(\ref{UjN}).

\section{Ising gate}

\label{ApIsing}

The controlled phase shift gate is decomposed by using the Z gate and the
Ising gate as%
\begin{equation}
\text{CP}\left( \phi _{j}\right) =e^{-i\frac{\phi _{j}}{4}}e^{i\frac{\phi
_{j}}{4}\text{Z}_{j}\otimes \text{I}_{j+1}}e^{i\frac{\phi _{j}}{4}\text{I}%
_{j}\otimes \text{Z}_{j+1}}e^{-i\frac{\phi _{j}}{4}\text{Z}_{j}\otimes \text{%
Z}_{j+1}}.
\end{equation}%
In actual experiments, the CZ gate and the CP gate are constructed by using
this formula. The product of the CP gate is%
\begin{equation}
\prod_{j=1}^{2L}\text{CP}\left( \phi _{j}\right) =\prod_{j=1}^{2L}e^{-i%
\frac{\phi _{j}}{4}}e^{i\frac{\left( \phi _{j-1}+\phi _{j}\right) }{4}\text{Z%
}_{j}}e^{-i\frac{\phi _{j}}{4}\text{Z}_{j}\otimes \text{Z}_{j+1}}
\end{equation}%
for a closed chain, and 
\begin{align}
& \prod_{j=1}^{2L-1}\text{CP}\left( \phi _{j}\right)  \notag \\
=& e^{i\frac{\phi _{1}}{4}\text{Z}_{1}}e^{i\frac{\phi _{2L}}{4}\text{Z}%
_{2L}}\prod_{j=2}^{2L-1}e^{-i\frac{\phi _{j}}{4}}e^{i\frac{\left( \phi
_{j-1}+\phi _{j}\right) }{4}\text{Z}_{j}}e^{-i\frac{\phi _{j}}{4}\text{Z}%
_{j}\otimes \text{Z}_{j+1}}  \label{CPphi}
\end{align}%
for an open chain.

In general, it is hard to tune the angles of the Z gate and the Ising gate.
By generalizing (\ref{CPphi}), it is natural to consider independent angles
for the Z gate and the ZZ Ising gate as 
\begin{equation}
\text{ZZ}\equiv \prod_{j=1}^{2L}e^{i\frac{\phi _{j}^{\text{Z}}}{4}\text{Z}%
_{j}}\prod_{j=1}^{2L-1}e^{-i\frac{\phi _{j}^{\text{ZZ}}}{4}\text{Z}%
_{j}\otimes \text{Z}_{j+1}}.
\end{equation}%
The cluster state is generated by 
\begin{equation}
\left\vert \text{C}\left( \left\{ \phi _{j}^{\text{Z}},\phi _{j}^{\text{ZZ}%
}\right\} \right) \right\rangle \equiv \text{ZZ}\bigotimes_{j=1}^{L}\left%
\vert +\right\rangle .
\end{equation}%
The corresponding Hamiltonian is given by%
\begin{equation}
\mathcal{H}_{\text{CP}}=\alpha _{\zeta _{j-1}\zeta _{j+1}}\text{Z}%
_{j-1}^{\zeta _{j-1}}\text{X}_{j}\text{Z}_{j+1}^{\zeta _{j+1}}+\beta
_{\upsilon _{j-1}\upsilon _{j+1}}\text{Y}_{j-1}^{\upsilon _{j-1}}\text{X}_{j}%
\text{Y}_{j+1}^{\upsilon _{j+1}}
\end{equation}%
with%
\begin{align}
\alpha _{00}=& 0, \\
\alpha _{10}=& \alpha _{01}=\frac{\sin ^{2}\phi _{j}^{\text{Z}}}{2}, \\
\alpha _{11}=& -\sin ^{2}\frac{\phi _{j}^{\text{Z}}}{2}\cos \phi _{j}, \\
\beta _{00}=& \cos ^{2}\frac{\phi _{j}}{2}\sin \phi _{j}, \\
\beta _{10}=& \beta _{01}=-\frac{\sin 2\phi _{j}}{4}, \\
\beta _{11}=& -\sin ^{2}\frac{\phi _{j}}{2}\sin \phi _{j}.
\end{align}

\subsection{Cross-resonance gate}

The cross-resonance gate is defined by\cite{CorcoS}%
\begin{align}
U_{\text{CR}}\equiv & e^{-i\frac{\phi _{j}^{\text{CR}}}{2}\text{Z}%
_{j}\otimes \text{X}_{j+1}}  \notag \\
=& \frac{1}{\sqrt{2}}\left( 
\begin{array}{cccc}
\cos \frac{\phi _{j}^{\text{CR}}}{2} & -i\sin \frac{\phi _{j}^{\text{CR}}}{2}
& 0 & 0 \\ 
-i\sin \frac{\phi _{j}^{\text{CR}}}{2} & \cos \frac{\phi _{j}^{\text{CR}}}{2}
& 0 & 0 \\ 
0 & 0 & \cos \frac{\phi _{j}^{\text{CR}}}{2} & i\sin \frac{\phi _{j}^{\text{%
CR}}}{2} \\ 
0 & 0 & i\sin \frac{\phi _{j}^{\text{CR}}}{2} & \cos \frac{\phi _{j}^{\text{%
CR}}}{2}%
\end{array}%
\right) ,
\end{align}%
which is naturally equipped in transmon-type superconducting qubits. By
applying the Hadamard gate to the $\left( j+1\right) $-th bit, we obtain the
Ising gate.

\subsection{M\o lmer-S\o rensen gate}

The M\o lmer-S\o rensen gate is defined by\cite{MSS}%
\begin{align}
U_{\text{MS}}\equiv & e^{-i\frac{\phi _{j}^{\text{MS}}}{2}\text{X}%
_{j}\otimes \text{X}_{j+1}}  \notag \\
=& \frac{1}{\sqrt{2}}\left( 
\begin{array}{cccc}
\cos \frac{\phi _{j}^{\text{MS}}}{2} & 0 & 0 & -i\sin \frac{\phi _{j}^{\text{%
MS}}}{2} \\ 
0 & \cos \frac{\phi _{j}^{\text{MS}}}{2} & -i\sin \frac{\phi _{j}^{\text{MS}}%
}{2} & 0 \\ 
0 & i\sin \frac{\phi _{j}^{\text{MS}}}{2} & \cos \frac{\phi _{j}^{\text{MS}}%
}{2} & 0 \\ 
i\sin \frac{\phi _{j}^{\text{MS}}}{2} & 0 & 0 & \cos \frac{\phi _{j}^{\text{%
MS}}}{2}%
\end{array}%
\right) ,
\end{align}%
which is naturally equipped in qubits made of ion trap. By applying the
Hadamard gate to the $j$-th and $\left( j+1\right) $-th bits, we obtain the
Ising gate.

\section{Non-invertible symmetry in unitary-mapped cluster model}

\label{ApNI}

The generators (\ref{etaZXZ}) of the $\mathbb{Z}_{2}^{\text{even}}\times 
\mathbb{Z}_{2}^{\text{odd}}$ symmetry is modified as%
\begin{equation}
\eta ^{\text{even}}\equiv V\left( \prod\limits_{j\in \text{even}}\text{X}%
_{j}\right) V^{-1},\quad \eta ^{\text{odd}}\equiv V\left(
\prod\limits_{j\in \text{odd}}\text{X}_{j}\right) V^{-1}.
\end{equation}%
The generator (\ref{DZXZ}) of the non-invertible symmetry is modified as%
\begin{equation}
\mathsf{D}_{U}=V\mathsf{D}_{\text{ZXZ}}V^{-1}.
\end{equation}%
By using the relations%
\begin{align}
V\mathsf{D}_{\text{ZXZ}}V^{-1}V\text{X}_{j}V^{-1}=& V\text{Z}_{j-1}\text{Z}%
_{j+1}V^{-1}V\mathsf{D}_{\text{ZXZ}}V^{-1}, \\
V\mathsf{D}_{\text{ZXZ}}\text{Z}_{j-1}V^{-1}V\text{Z}_{j+1}V^{-1}=& V\text{X}%
_{j}V^{-1}V\mathsf{D}_{\text{ZXZ}}V^{-1},
\end{align}%
we obtain the action of the non-invertible symmetry%
\begin{align}
V\text{X}_{j}V^{-1}\rightsquigarrow & V\text{Z}_{j-1}\text{Z}_{j+1}V^{-1}, \\
V\text{Z}_{j-1}\text{Z}_{j+1}V^{-1}\rightsquigarrow & V\text{X}_{j}V^{-1}.
\end{align}%
Similarly, the action (\ref{KTZXZ}) of the KT transformation is modified as%
\begin{align}
V\text{X}_{j}V^{-1}\leadsto & V\text{X}_{j}V^{-1}, \\
V\text{Z}_{j-1}\text{X}_{j}\text{Z}_{j+1}V^{-1}\leadsto & V\text{Z}_{j-1}%
\text{Z}_{j+1}V^{-1}.
\end{align}%
The string-order parameter is modified as%
\begin{equation}
\mathcal{O}^{U}\left( i,j\right) \equiv \left\langle g_{U}\right\vert
\prod\limits_{k=i}^{j}K_{k}^{U}\left\vert g_{U}\right\rangle .
\end{equation}%

\section{Clifford hierarchy}

\label{ApCl}

We review the Clifford hierarchy\cite{Gottes}. The Pauli group belongs to
the Clifford hierarchy $\mathcal{C}_{1}$ with the level 1. The Clifford
hierarchy $\mathcal{C}_{k}$ is defined by the set which maps the Pauli group
within $\mathcal{C}_{k-1}$ as 
\begin{equation}
\mathcal{C}_{k}=\left\{ U|UPU^{\dagger }\in \mathcal{C}_{k-1},\forall P\in
P\right\} .
\end{equation}%
Explicit examples are%
\begin{align}
\mathcal{C}_{1}=& \left\{ \text{I},\text{X},\text{Y},\text{Z}\right\} , \\
\mathcal{C}_{2}=& \left\{ \text{H},\text{S},\text{CNOT},\text{CZ}\right\} ,
\\
\mathcal{C}_{3}=& \left\{ \text{T},\text{CCZ},\text{CS},\text{Toffol}%
i\right\} , \\
\mathcal{C}_{4}=& \left\{ \text{C}^{3}\text{Z},\text{C}^{2}\text{S},\text{CT}%
\right\} , \\
& \cdots \\
\mathcal{C}_{k}=& \left\{ \text{C}^{k-1}\text{Z},\text{C}^{k-2}\text{S},%
\text{C}^{k-3}\text{T}\right\} .
\end{align}%
C$^{N}$Z gate belong to the Clifford hierarchy with level ($N+1$). C$^{N}$S
gate belong to the Clifford hierarchy with level ($N+2$). The diagonal phase
shift gate%
\begin{equation}
R_{N}\equiv \text{diag.}\left( 1,e^{i\pi /2^{N-1}}\right)
\end{equation}%
belongs the Clifford hierarchy\cite{Cui} with the level $N$. Especially, $%
R_{1}=$Z, $R_{2}=$S and $R_{3}=$T. Then C$^{k}R_{N}$ gate belongs to the
Clifford hierarchy with the level $N+k$.

\end{document}